\documentclass[11pt]{article}
\pdfoutput=1

\usepackage{color}
\usepackage{mhchem}
\usepackage{xcolor}
\usepackage{colortbl}
\usepackage{hhline}
\usepackage{cite}
\usepackage{hyperref}
\hypersetup{colorlinks=true,linkcolor=red,anchorcolor=black,citecolor=green}
\usepackage[toc,page]{appendix}
\usepackage{mathtools}
\usepackage{amsfonts}
\usepackage{bbold}
\usepackage{textcomp}
\usepackage[DIV13]{typearea}
\usepackage{amsmath, amsthm, amssymb, mathtools,empheq,latexsym,dsfont}
\usepackage{bbm}
\usepackage{amsmath,amssymb,latexsym,dsfont}
\usepackage{slashed, simplewick}
\usepackage[utf8]{inputenc}
\usepackage{graphicx}
\usepackage{graphicx,placeins}
\usepackage{makeidx}
\usepackage[font=small,labelfont=bf]{caption}
\usepackage{nicefrac}
\usepackage{subfigure}
\usepackage{array, bigdelim,multirow,multicol}
\usepackage{mathrsfs}
\allowdisplaybreaks

\usepackage{adjustbox}
\usepackage{booktabs}
\usepackage{indentfirst}
\usepackage{booktabs}
\usepackage{tabulary}
\usepackage{longtable}
\usepackage{fancybox}
\usepackage{graphicx}
\usepackage{bm}
\usepackage{bbm}
\usepackage{tikz}
\usepackage{diagbox}
\usepackage{enumitem}
\usepackage[integrals]{wasysym}

\graphicspath{{immagini/}}

\definecolor{Gray}{gray}{0.92}

\newcommand{\ignore}[1]{}

\renewcommand{\arg}{{\rm Arg}}

\newcommand{\be}{\begin{equation}}
	\newcommand{\ee}{\end{equation}}
\newcommand{\bea}{\begin{eqnarray}}
	\newcommand{\eea}{\end{eqnarray}}

\definecolor{lightred}{rgb}{1,0.4,0.4}
\definecolor{lightgreen}{rgb}{0.4,1,0.4}
\definecolor{LightCyan}{rgb}{0.88,1,1}

\newcounter{Thm}[section]
\renewcommand{\theThm}{\arabic{section}.\arabic{Thm}}

\newcounter{nodecount}

\tikzstyle{every picture}+=[remember picture,baseline]
\tikzstyle{every node}+=[inner sep=0pt,anchor=base,
text depth=.25ex,outer sep=1.5pt]
\tikzstyle{every path}+=[thick, rounded corners]
\tikzset{
	plabel/.style={inner sep=2pt}
}

\makeatletter
\@addtoreset{equation}{section}
\makeatother

\begin{document}
  \title{\begin{center}
	{\Large\bf Minimal lepton models with non-holomorphic modular $A_{4}$ symmetry}
\end{center}}
    \date{}
	\author{Xiang-Yan Gao$^{a,b,c}$\footnote{E-mail: {\tt 202421359@stumail.nwu.edu.cn}},  \
		Cai-Chang Li$^{a,b,c}$\footnote{E-mail: {\tt ccli@nwu.edu.cn}} \ \\*[20pt]
		\centerline{\begin{minipage}{\linewidth}
				\begin{center}
					$^a${\it\small School of Physics, Northwest University, Xi'an 710127, China}\\[2mm]
					$^b${\it\small Shaanxi Key Laboratory for Theoretical Physics Frontiers, Xi'an 710127, China}\\[2mm]
					$^c${\it\small NSFC-SPTP Peng Huanwu Center for Fundamental Theory, Xi'an 710127, China}\\[2mm]
				\end{center}
\end{minipage}} \\[10mm]}
	\maketitle
	
	\thispagestyle{empty}

\centerline{\large\bf Abstract}
\begin{quote}
We present a comprehensive bottom-up analysis of lepton mass and mixing based on the non-holomorphic $A_{4}$ modular symmetry. Neutrinos are assumed to be Majorana particles and the light neutrino masses are generated through the Weinberg operator. In this framework, we construct all phenomenologically viable models with minimal number of free parameters, where the Yukawa couplings are expressed in terms of polyharmonic Maa{\ss} forms of weights $\pm4$, $\pm2$ and $0$ at level $N=3$. Without imposing generalized CP (gCP) symmetry, we identify 147 (6) viable models with seven real free parameters that successfully reproduce the current experimental data of lepton sector for the normal (inverted) mass ordering.  When gCP symmetry consistent with  $A_{4}$ modular symmetry is included, the number of free parameters is reduced by one, yielding 47 (5) phenomenologically viable models in the normal (inverted) mass ordering. Finally, we present detailed numerical analyses of a representative model for both mass orderings to illustrate these results.

\end{quote}
	\newpage

\section{Introduction}	

The Standard Model (SM) successfully describes particle interactions but leaves unexplained the origin of fermion families and their highly hierarchical masses and mixing patterns—a challenge known as the flavor problem~\cite{ParticleDataGroup:2024cfk}. The masses of fundamental particles span more than twelve orders of magnitude, ranging from the neutrino mass on the order of a fraction of an eV to the top quark mass of 173 GeV. 
Moreover, the quark mixing angles are small, whereas the lepton mixing angles consist of two large angles and one small angle comparable in magnitude to the Cabibbo angle~\cite{ParticleDataGroup:2024cfk}. The SM attributes these patterns to arbitrary Yukawa couplings, providing no underlying principle for their observed hierarchy and structure.

Several approaches have been developed to address the flavor problem. One line of research introduces flavor symmetries, particularly non-Abelian discrete groups such as $A_{4}$, $S_{4}$ and $A_{5}$ which can naturally reproduce large lepton mixing\cite{Altarelli:2010gt,Ishimori:2010au,King:2013eh,King:2014nza,King:2015aea,King:2017guk,Petcov:2017ggy,Xing:2020ijf,Feruglio:2019ybq,Almumin:2022rml,Ding:2024ozt}. In this framework, the spontaneous breaking of such a family symmetry by scalar flavon VEVs gives rise to the vacuum alignment problem. This, in turn, introduces further complications. Modular invariance addresses these issues by offering an economical framework in which Yukawa couplings are modular forms transforming as irreducible representations of the finite modular groups $\Gamma_{N}$ or $\Gamma^{\prime}_{N}$, thereby eliminating flavons and removing the need for vacuum alignment~\cite{Feruglio:2017spp,Kobayashi:2023zzc,Ding:2023htn}. 
This approach enables highly predictive fermion mass models. In the minimal modular invariant model, all lepton masses and mixing parameters are determined by only four real parameters plus the modulus $\tau$~\cite{Ding:2022nzn,Ding:2023ydy}. The same modulus can also link quark and lepton sectors, allowing a simultaneous description of their flavor observables with just fourteen real parameters, including the real and imaginary parts of $\tau$~\cite{Ding:2023ydy,Ding:2024pix}.
However, the predictive power of this framework is limited because modular symmetry only weakly constrains the K$\ddot{\mathrm{a}}$hler potential~\cite{Chen:2019ewa}. Integrating modular symmetry with a traditional flavor symmetry resolves this limitation and gives rise to either an eclectic flavor group~\cite{Baur:2019kwi,Baur:2019iai,Nilles:2020tdp,Nilles:2020gvu,Nilles:2020nnc,Nilles:2020kgo,Ding:2023ynd,Li:2023dvm,Li:2024pff} or quasi-eclectic flavor symmetry~\cite{Chen:2021prl}.

In the framework of the original modular invariance approach, the Yukawa couplings are assumed to be modular forms of level $N$ which are holomorphic functions of the complex modulus $\tau$. To preserve this holomorphicity, supersymmetry (SUSY) is required~\cite{Lauer:1989ax,Ferrara:1989bc,Ferrara:1989qb,Feruglio:2017spp}. However, experimental evidence for low-energy SUSY remains elusive, making its natural realization uncertain. This motivates the study of non-supersymmetric modular invariant theories. Recently, a non-holomorphic modular flavor symmetry framework has been developed, which remains valid in non-supersymmetric settings~\cite{Qu:2024rns}. In this framework, the holomorphicity condition is replaced by a harmonic one, and the Yukawa couplings are required to be polyharmonic Maa{\ss} forms of level $N$ and even weight. These forms can be decomposed into multiplets of the inhomogeneous finite modular group $\Gamma_{N}$. Unlike holomorphic modular forms, which are defined only for non-negative weights $k\geq0$, the polyharmonic Maa{\ss} forms extend to negative weights. Moreover, non-holomorphic polyharmonic Maa{\ss} exist for weights $k=0,1$ and $2$. Phenomenologically viable models based on finite modular groups such as $\Gamma_2\cong S_3$~\cite{Okada:2025jjo}, $\Gamma_3\cong A_4$~\cite{Qu:2024rns,Kumar:2024uxn,Nomura:2024atp,Nomura:2024nwh,Kobayashi:2025hnc,Loualidi:2025tgw,Kumar:2025bfe,Nomura:2025ovm,Nomura:2025raf,Zhang:2025dsa,Priya:2025wdm,Kumar:2025nut,Nanda:2025lem,Jangid:2025thp},  $\Gamma_4\cong S_4$~\cite{Ding:2024inn}, and  $\Gamma_5\cong A_5$~\cite{Li:2024svh} have been successfully constructed. The framework has further been extended to include odd-weight polyharmonic Maa{\ss} forms, which transform under irreducible representations of  the homogeneous finite modular groups $\Gamma^{\prime}_{N}$~\cite{Qu:2025ddz}. Modular invariant models based on the corresponding non-holomorphic groups $\Gamma^{\prime}_{3}\cong T^{\prime}$~\cite{Qu:2025ddz} and $\Gamma^{\prime}_{5}\cong A^{\prime}_{5}$~\cite{Li:2025kcr} have also been studied. Furthermore, the non-holomorphic modular symmetry can consistently combine with generalized CP (gCP) symmetry, which restricts the phases of couplings, boosting modular invariant model predictions, as in supersymmetric modular flavor symmetry~\cite{Novichkov:2019sqv}. Under modular transformations, the gCP operation acts as $\tau \stackrel{{\rm CP}}{\longmapsto} -\tau^*$~\cite{Dent:2001cc,Dent:2001mn,Baur:2019kwi,Baur:2019iai,Novichkov:2019sqv}. In a basis with unitary and symmetric $S$ and $T$ matrices across all irreducible representations, this gCP transformation reduces to conventional CP, represented by the identity in flavor space.

This work provides a systematic analysis of lepton models based on non-holomorphic $A_{4}$ modular symmetry. 
We focus on the most economical modular invariant constructions, which require no flavon fields other than the modulus $\tau$ and generate light neutrino masses through the Weinberg operator.
The three generations of left-handed leptons are assigned to the irreducible triplet representation $\bm{3}$ of $A_{4}$, while the right-handed leptons are allowed to transform as all possible combinations of the singlet representations $\bm{1}$, $\bm{1^{\prime}}$ and $\bm{1^{\prime\prime}}$. Using the level 3 polyharmonic Maa{\ss} forms of weights $k=\pm4,\, \pm2,\, 0$, we construct all phenomenologically viable and minimally parameterised lepton models. Without imposing gCP symmetry, 147 models for normal ordering (NO) and 6 models for inverted ordering (IO) successfully reproduce the experimental data. When gCP symmetry is enforced, 47 of the 147 NO models and 5 of the 6 IO models remain consistent with the observations in the lepton sector. 
The current JUNO~\cite{JUNO:2025gmd}  constraint on $\sin^{2}\theta_{12}$ rules out only 5 of the 147 viable NO models without gCP symmetry. All others remain consistent with current measurements. Notably, the non-holomorphic $A_4$ modular symmetry yields a richer set of level $3$ polyharmonic Maa{\ss} forms than its supersymmetric framework~\cite{Ding:2019zxk}. This expanded landscape allows the construction of more viable minimal lepton models consistent with experimental data.

We organise the rest of this paper in the following.  In section~\ref{sec:NHMF}, we briefly review the framework of non-holomorphic modular flavor symmetry at level $N=3$. Section~\ref{sec:model} presents a class of minimal lepton flavor models invariant under the non-holomorphic $A_4$ modular symmetry, and examines their predictions for lepton mixing angles, CP-violating phases and neutrino masses. A representative model is introduced in section~\ref{sec:example-models}, where a detailed numerical analysis is performed for both NO and IO neutrino mass spectra.  We conclude the paper in section~\ref{sec:conclusion}.

\section{\label{sec:NHMF}Non-holomorphic modular flavor symmetry of level $N=3$}

This section provides a concise overview of non-holomorphic modular flavor symmetry. The modular group $\Gamma \equiv SL(2, \mathbb{Z})$  consists of $2 \times 2$ integer matrices with determinant one. Its action on the upper half complex plane $\mathcal{H}=\left\{\tau\in\mathbb{C}|\Im(\tau)>0\right\}$ via fractional linear transformations:
\begin{equation}
 \tau\mapsto\gamma\tau=\frac{a\tau+b}{c\tau+d}\, , \qquad    \gamma=\begin{pmatrix}
		a & b \\
		c & d
	\end{pmatrix}\in SL(2,\mathbb{Z})\,,
\end{equation}
where the modular group $SL(2,\mathbb{Z})$ can be generated by two matrices $S$ and $T$, which are given by
\begin{equation}
	S=\begin{pmatrix}
		0 & 1 \\
		-1 & 0
	\end{pmatrix}, ~~~~
	T=\begin{pmatrix}
		1 & 1 \\
		0 & 1
	\end{pmatrix}\,,
\end{equation}
which act on $\tau$ as:
\begin{equation}
	\tau\mapsto\ S\tau=-\frac{1}{\tau}\,,\qquad \tau\mapsto\ T\tau=\tau+1\,.
\end{equation}
Note that $\gamma$ and $-\gamma$ induce the same transformation on $\tau$. The finite modular group $\Gamma_N$ is obtained via the quotient:
\begin{equation}\label{eq:Gamma_N_Def}
\Gamma_N\equiv \Gamma/\pm\Gamma(N)\,,
\end{equation}
where  the principal congruence subgroup $\Gamma(N)$ of level $N$ for any positive integer $N$ is defined as:
\begin{equation}
  \Gamma(N)=\left\{
  \begin{pmatrix}
	a & b \\
	c & d
  \end{pmatrix} \in SL(2,\mathbb{Z}),\quad
  \begin{pmatrix}
    a & b \\
    c & d
  \end{pmatrix}\equiv
  \begin{pmatrix}
	1 & 0 \\
	0 & 1
  \end{pmatrix}(\text{mod}~ N)\right\}\,,
\end{equation}
which is a normal subgroup of the special linear group $\mathrm{SL}(2,\mathbb{Z})$. For $N = 2, 3, 4$ and $5$, the finite modular group $\Gamma_N$ is isomorphic to the groups $S_3$, $A_4$, $S_4$ and $A_5$, respectively~\cite{deAdelhartToorop:2011re}. These finite groups are generated by the modular transformations $S$ and $T$, subject to the relations:
\begin{equation}
	S^2=(ST)^3=T^{N}=1\,.
\end{equation}
Additionally, the quotient $\Gamma^{\prime}_N \equiv \Gamma/\Gamma(N)$ forms a double cover of $\Gamma_N$~\cite{Liu:2019khw}.

In the present work, we are interested in the finite modular group $\Gamma_{3}\cong A_{4}$ which can be seen as the symmetry group of the tetrahedron. The $A_{4}$ group has four inequivalent irreducible representations: three singlet representations $\bm{1}$, $\bm{1^{\prime}}$, $\bm{1^{\prime\prime}}$, and one triplet representation $\bm{3}$.  The generators $S$ and $T$ in these irreducible representations can be expressed as:
\begin{equation}
  \begin{array}{rll}
	\bm{1:} &   ~S=1\,,~ &  T=1 \,, ~\\[-14pt] \\[4pt]
	\bm{1^{\prime}:} &   ~S=1\,,~ &  T=\omega \,, ~\\[-14pt] \\[4pt]
	\bm{1^{\prime\prime}:} & ~S=1\,,~ & T=\omega^2 \,, ~\\[-14pt] \\[4pt]
	\bm{3:} &   ~S=\frac{1}{3}
		\begin{pmatrix}
			-1 &  2 & 2 \\
			 2 & -1 & 2 \\
			 2 &  2 & -1
		\end{pmatrix}\,,~
		& T=\begin{pmatrix}
			1 & 0 & 0 \\
			0 & \omega & 0  \\
			0 & 0  & \omega^2
		    \end{pmatrix}\,, ~~
  \end{array}
\end{equation}
with $\omega=e^{2\pi i/3}$. The Kronecker products between any two of the four irreducible representations of $A_{4}$ are given by:
\begin{eqnarray}
	\nonumber&&\bm{1}\otimes\bm{R}=\bm{R}\otimes\bm{1}=\bm{R},\qquad \bm{1^{\prime}}\otimes\bm{1^{\prime}}=\bm{1^{\prime\prime}}, \qquad \bm{1^{\prime}}\otimes\bm{1^{\prime\prime}}=\bm{1}, \qquad \bm{1^{\prime\prime}}\otimes\bm{1^{\prime\prime}}=\bm{1^{\prime}}\\
    \label{eq:Kronecker}&&\bm{1^{\prime}}\otimes\bm{3}=\bm{3},\qquad \bm{1^{\prime\prime}}\otimes\bm{3}=\bm{3}, \qquad \bm{3}\otimes\bm{3}=\bm{1}\oplus\bm{1^{\prime}}\oplus\bm{1^{\prime\prime}}\oplus\bm{3_{S}}\oplus\bm{3_{A}},
\end{eqnarray}
where $\bm{R}$  denotes any irreducible representation of $A_{4}$, and  $\bm{3_{S}}$, $\bm{3_{A}}$  refer to the symmetric and antisymmetric combinations, respectively. For any two triplets $\alpha=(\alpha_{1},\alpha_{2},\alpha_{3})$ and $\beta=(\beta_{1},\beta_{2},\beta_{3})$, the decomposition of the tensor product of them is ~\cite{Altarelli:2005yx}:
\begin{eqnarray}
\nonumber&&(\alpha\beta)_{\bm{1}}=\alpha_{1}\beta_{1}+\alpha_{2}\beta_{3}+\alpha_{3}\beta_{2},\\
\nonumber&&(\alpha\beta)_{\bm{1}^{\prime}}=\alpha_{3}\beta_{3}+\alpha_{1}\beta_{2}+\alpha_{2}\beta_{1},\\
\nonumber&&(\alpha\beta)_{\bm{1}^{\prime\prime}}=\alpha_{2}\beta_{2}+\alpha_{1}\beta_{3}+\alpha_{3}\beta_{1},\\
\nonumber&&(\alpha\beta)_{\bm{3_{S}}}=(2\alpha_{1}\beta_{1}-\alpha_{2}\beta_{3}-\alpha_{3}\beta_{2},2\alpha_{3}\beta_{3}-\alpha_{1}\beta_{2}-\alpha_{2}\beta_{1},2\alpha_{2}\beta_{2}-\alpha_{1}\beta_{3}-\alpha_{3}\beta_{1}),\\
\label{eq:Kronecker}&&(\alpha\beta)_{\bm{3_{A}}}=(\alpha_{2}\beta_{3}-\alpha_{3}\beta_{2},\alpha_{1}\beta_{2}-\alpha_{2}\beta_{1},\alpha_{3}\beta_{1}-\alpha_{1}\beta_{3}).
\end{eqnarray}

In the framework of non-holomorphic modular flavor symmetry, the Lagrangian is built from matter fields together with polyharmonic Maa{\ss} forms of level $N$, which are functions of the complex modulus $\tau$ and can include both holomorphic and non-holomorphic parts. These forms, of even weight $k$  and level $N$, constitute a finite dimensional linear space and can be arranged into modular multiplets $Y^{(k)}_{\bm{r}}(\tau)$, which transform under irreducible representations $\bm{r}$ of  $\Gamma_{N}$ up to an automorphy factor:
\begin{equation}
Y^{(k)}_{\bm{r}}(\tau)\mapsto Y^{(k)}_{\bm{r}}(\gamma\tau)=(c\tau+d)^{k}\rho_{Y}(\gamma)Y^{(k)}_{\bm{r}}(\tau)\,.
\end{equation}
In this work, we present a comprehensive analysis of  all the simplest lepton models based on non-holomorphic $A_{4}$ modular flavor symmetry,  using the level $3$ polyharmonic Maa{\ss} forms with weights  $k = \pm4,\, \pm2,\, 0$, as summarized in table~\ref{tab:MF_summary}. The explicit expressions for these forms are omitted here due to their length, and they can be found in Ref.~\cite{Qu:2024rns}.

\begin{table}[t!]
	\centering
	\renewcommand{\arraystretch}{1.3}
	\begin{tabular}{|c|c|c|c|c|c|}
		\hline  \hline
		
		Weight $k_Y$ &  $k_Y=-4$  & 	$k_Y=-2$ &  $k_Y=0$ & $k_Y=2$ & $k_Y=4$  \\ \hline
		
	$Y^{(k_Y)}_{\bm{r}}$	 & $Y^{(-4)}_{\bm{1}}$,\; $Y^{(-4)}_{\bm{3}}$ & $Y^{(-2)}_{\bm{1}}$,\; $Y^{(-2)}_{\bm{3}}$ & $Y^{(0)}_{\bm{1}}$,\; $Y^{(0)}_{\bm{3}}$ & $Y^{(2)}_{\bm{1}}$,\; $Y^{(2)}_{\bm{3}}$ & $Y^{(4)}_{\bm{1}}$,\; $Y^{(4)}_{\bm{1}'}$,\; $Y^{(4)}_{\bm{3}}$ \\ \hline \hline

	\end{tabular}
	\caption{\label{tab:MF_summary}Polyharmonic Maa{\ss} form multiplets of level $3$ and weights $k=\pm4,\,\pm2,\,0$, the subscript $\bm{r}$ denotes the transformation property under finite group $A_{4}$. The explicit forms of these polyharmonic  Maa{\ss} form multiplets can be found in Ref.~\cite{Qu:2024rns}.} 
\end{table}

Additionally, the gCP symmetry can be consistently included in the framework of non-holomorphic modular symmetry~\cite{Qu:2024rns,Qu:2025ddz}.  It turns out that the gCP invariance enforces the coupling constants accompanying each invariant singlet in the Lagrangian to be real in the basis where both modular generators $S$ and $T$ are represented by unitary and symmetric matrices. In the basis adopted here, $T$ is represented by diagonal matrices in different $A_{4}$ irreducible representations, and the representation matrices of $S$ are real and symmetric. As a result, gCP symmetry simply requires all couplings to be real in our models.

\section{\label{sec:model}Lepton models based on non-holomorphic $A_{4}$ modular symmetry }

In this section, we systematically classify the minimal lepton mass models based on finite modular symmetry $\Gamma_{3}\cong A_{4}$ in the framework of non-supersymmetry. The Yukawa couplings are described by the polyharmonic Maa{\ss} forms of level $N=3$ and  even weights, which can be decomposed  into the irreducible multiplets of  $A_{4}$. The polyharmonic Maa{\ss} form multiplets of level $3$ and weights $\pm 4$, $\pm2$ and $0$ are listed in table~\ref{tab:MF_summary}. In the present work, neutrinos are assumed to be Majorana particles, and the light neutrino masses are generated via the Weinberg operator. The analysis focuses on the most economical scenarios, in which modular invariance is attained without introducing any flavon fields other than the complex modulus $\tau$. 
We assume that the Higgs doublet fields $H$ have vanishing modular weight and transform as the trivial singlet $\bm{1}$ of $A_{4}$. The left-handed (LH) lepton doublets $L$ with modular weight $k_L$ form a triplet $\bm{3}$, while the three right-handed (RH) charged leptons $E^c_{1,2,3}$ are $A_4$ singlets with modular weights $k_{E^{C}_{1,2,3}}$. 
To minimize the number of free parameters, we restrict our analysis to polyharmonic  Maa{\ss}  forms of level 3 with even weights ranging from $k =-4$ to $4$, prioritizing the use of the lowest weight forms whenever possible.
	
\subsection{\label{sec:ch_mass}Charged lepton sector }

We find ten distinct independent representation assignments for the lepton fields, with the LH doublets forming a triplet and the RH charged leptons as singlets. Among these, three assignments feature all three RH leptons transforming under the same singlet, six have two transforming under one singlet and the third under another, and one has all three transforming under distinct singlets. The charged lepton sector is therefore divided into ten different cases, labeled   $C^{(k_{1},k_{2},k_{3})}_{1}$ to  $C^{(k_{1},k_{2},k_{3})}_{10}$, with the corresponding representation assignments summarized in table~\ref{tab:sum_ch}. For all ten assignments, the most general Lagrangian $\mathcal{L}_e$ describing the charged lepton masses are given by:
\begin{eqnarray}
\nonumber C^{(k_{1},k_{2},k_{3})}_{1}&:&-\mathcal{L}_e=\left[\alpha E^{c}_1\left(LY^{(k_{1})}_{\bm{3}}\right)_{\bm{1}}+
	\beta E^{c}_2\left(LY^{(k_{2})}_{\bm{3}}\right)_{\bm{1}}+\gamma E^{c}_3\left(LY^{(k_{3})}_{\bm{3}}\right)_{\bm{1}}\right]H^{*}\,, \\
\nonumber C^{(k_{1},k_{2},k_{3})}_{2}&:&-\mathcal{L}_e=\left[\alpha E^{c}_1\left(LY^{(k_{1})}_{\bm{3}}\right)_{\bm{1^{\prime\prime}}}+
	\beta E^{c}_2\left(LY^{(k_{2})}_{\bm{3}}\right)_{\bm{1^{\prime\prime}}}+\gamma E^{c}_3\left(LY^{(k_{3})}_{\bm{3}}\right)_{\bm{1^{\prime\prime}}}\right]H^{*}\,, \\
\nonumber C^{(k_{1},k_{2},k_{3})}_{3}&:&-\mathcal{L}_e=\left[\alpha E^{c}_1\left(LY^{(k_{1})}_{\bm{3}}\right)_{\bm{1^{\prime}}}+
	\beta E^{c}_2\left(LY^{(k_{2})}_{\bm{3}}\right)_{\bm{1^{\prime}}}+\gamma E^{c}_3\left(LY^{(k_{3})}_{\bm{3}}\right)_{\bm{1^{\prime}}}\right]H^{*}\,, \\
\nonumber C^{(k_{1},k_{2},k_{3})}_{4}&:&-\mathcal{L}_e=\left[\alpha E^{c}_1\left(LY^{(k_{1})}_{\bm{3}}\right)_{\bm{1}}+
	\beta E^{c}_2\left(LY^{(k_{2})}_{\bm{3}}\right)_{\bm{1}}+\gamma E^{c}_3\left(LY^{(k_{3})}_{\bm{3}}\right)_{\bm{1^{\prime\prime}}}\right]H^{*}\,, \\
\nonumber C^{(k_{1},k_{2},k_{3})}_{5}&:&-\mathcal{L}_e=\left[\alpha E^{c}_1\left(LY^{(k_{1})}_{\bm{3}}\right)_{\bm{1}}+
	\beta E^{c}_2\left(LY^{(k_{2})}_{\bm{3}}\right)_{\bm{1}}+\gamma E^{c}_3\left(LY^{(k_{3})}_{\bm{3}}\right)_{\bm{1^{\prime}}}\right]H^{*}\,, \\
\nonumber C^{(k_{1},k_{2},k_{3})}_{6}&:&-\mathcal{L}_e=\left[\alpha E^{c}_1\left(LY^{(k_{1})}_{\bm{3}}\right)_{\bm{1^{\prime\prime}}}+
	\beta E^{c}_2\left(LY^{(k_{2})}_{\bm{3}}\right)_{\bm{1^{\prime\prime}}}+\gamma E^{c}_3\left(LY^{(k_{3})}_{\bm{3}}\right)_{\bm{1}}\right]H^{*}\,, \\
\nonumber C^{(k_{1},k_{2},k_{3})}_{7}&:&-\mathcal{L}_e=\left[\alpha E^{c}_1\left(LY^{(k_{1})}_{\bm{3}}\right)_{\bm{1^{\prime\prime}}}+
	\beta E^{c}_2\left(LY^{(k_{2})}_{\bm{3}}\right)_{\bm{1^{\prime\prime}}}+\gamma E^{c}_3\left(LY^{(k_{3})}_{\bm{3}}\right)_{\bm{1^{\prime}}}\right]H^{*}\,, \\
\nonumber C^{(k_{1},k_{2},k_{3})}_{8}&:&-\mathcal{L}_e=\left[\alpha E^{c}_1\left(LY^{(k_{1})}_{\bm{3}}\right)_{\bm{1^{\prime}}}+
	\beta E^{c}_2\left(LY^{(k_{2})}_{\bm{3}}\right)_{\bm{1^{\prime}}}+\gamma E^{c}_3\left(LY^{(k_{3})}_{\bm{3}}\right)_{\bm{1}}\right]H^{*}\,, \\
\nonumber C^{(k_{1},k_{2},k_{3})}_{9}&:&-\mathcal{L}_e=\left[\alpha E^{c}_1\left(LY^{(k_{1})}_{\bm{3}}\right)_{\bm{1^{\prime}}}+
	\beta E^{c}_2\left(LY^{(k_{2})}_{\bm{3}}\right)_{\bm{1^{\prime}}}+\gamma E^{c}_3\left(LY^{(k_{3})}_{\bm{3}}\right)_{\bm{1^{\prime\prime}}}\right]H^{*}\,, \\
\label{eq:Le_10case} C^{(k_{1},k_{2},k_{3})}_{10}&:&-\mathcal{L}_e=\left[\alpha E^{c}_1\left(LY^{(k_{1})}_{\bm{3}}\right)_{\bm{1}}+
	\beta E^{c}_2\left(LY^{(k_{2})}_{\bm{3}}\right)_{\bm{1^{\prime\prime}}}+\gamma E^{c}_3\left(LY^{(k_{3})}_{\bm{3}}\right)_{\bm{1^{\prime}}}\right]H^{*}\,,
\end{eqnarray}
where the  weights of polyharmonic Maa{\ss} form triplets $Y^{(k_{i})}_{\bm{3}}$ satisfy $k_{i}=k_{L}+k_{E^{c}_{i}}$ ($i=1,2,3$), and  the phases of the couplings $\alpha$, $\beta$, and $\gamma$ can be fully absorbed by rephasing the RH charged lepton fields $E^{c}_{1}$, $E^{c}_{2}$, and $E^{c}_{3}$, respectively. As a result, these couplings can be taken as real and positive. 

In each of the first three cases in Eq.~\eqref{eq:Le_10case}, all three RH charged leptons $E^{c}_{i}$ transform under the same $A_4$ singlet representation. They are distinguished by their modular weights $k_{E^{c}_{i}}$, which are achieved by coupling each to modular form multiplets of different weights.  In this work, the charged lepton mass matrix $M_e$ is defined by the convention $E^{c}M_{e}L$. Permuting any two rows of $M_e$ corresponds to a field redefinition of the corresponding RH leptons and leaves predictions for masses and mixing parameters unchanged. Similarly, exchanging the modular weights of any two RH charged lepton fields also yields identical physical predictions. For minimality and simplicity, we assume $k_1  < k_2  < k_3  \in \{\pm4, \pm2, 0\}$ without loss of generality, which results in 10 distinct charged lepton mass matrices from the 10 independent weight assignments in each of the three cases. The corresponding lepton mass matrices are summarized in table~\ref{tab:sum_ch}.

For the six cases $C^{(k_1,k_2,k_3)}_4$ to $C^{(k_1,k_2,k_3)}_9$, two RH charged leptons are assigned to one $A_{4}$ singlet, while the third is assigned to a different singlet. Without loss of generality, we assume that $E^{c}_1$ and $E^{c}_2$ transform under the same singlet, and $E^{c}_3$ under another. This leads to 50 independent weight assignments for each of the six cases, obtained from the combinations satisfying: $k_{1} < k_{2} \in \{\pm4, \pm2, 0\}$ and $k_{3} \in \{\pm4, \pm2, 0\}$. The corresponding representation assignments to RH charged lepton fields, along with the charged lepton mass matrices are presented table~\ref{tab:sum_ch}.

In the final case $C^{(k_1,k_2,k_3)}_{10}$, the three RH charged leptons $E^{c}_{i}$ are assigned to the three distinct $A_{4}$ singlets $\bm{1}$, $\bm{1^{\prime}}$ and $\bm{1^{\prime\prime}}$. Their modular weights may be identical. For minimality and simplicity, the weights $k_i = k_L + k_{E^c_i}$ are chosen from the set ${\pm4, \pm2, 0}$, resulting in a total of 125 possible weight assignments for $k_i$. The charged lepton mass matrix for each weight assignment is given in table~\ref{tab:sum_ch}.

\begin{table}[t!]
\begin{center}
\renewcommand{\arraystretch}{1.31}
\begin{tabular}{|c|c|c|c|c|c|}  \hline\hline
\texttt{Cases} & $(\rho_{E^{c}_{1}},\rho_{E^{c}_{2}},\rho_{E^{c}_{3}})$  & $(k_{{1}},k_{{2}},k_{{3}})$ & $M_{e}$ \\  \hline
$C^{(k_{1},k_{2},k_{3})}_{1}$ & $(\bm{1},\bm{1},\bm{1})$ & & $\begin{pmatrix}\alpha Y^{(k_{1})}_{\bm{3},1}&\alpha Y^{(k_{1})}_{\bm{3},3}&\alpha Y^{(k_{1})}_{\bm{3},2}\\
	\beta Y^{(k_{2})}_{\bm{3},1}&\beta Y^{(k_{2})}_{\bm{3},3}&\beta Y^{(k_{2})}_{\bm{3},2}\\
	\gamma Y^{(k_{3})}_{\bm{3},1}&\gamma Y^{(k_{2})}_{\bm{3},3}&\gamma Y^{(k_{3})}_{\bm{3},2}\end{pmatrix}v$    \\ \cline{1-2} \cline{4-4} 

  $C^{(k_{1},k_{2},k_{3})}_{2}$ & $(\bm{1^{\prime}},\bm{1^{\prime}},\bm{1^{\prime}})$ & $k_{{1}}<k_{{2}}<k_{{3}}\in\{\pm4,\pm2,0\}$ & $\begin{pmatrix}{\alpha Y^{(k_{1})}_{\bm{3},3}}&{\alpha Y^{(k_{1})}_{\bm{3},2}}&{\alpha Y^{(k_{1})}_{\bm{3},1}}\\
	{\beta Y^{(k_{2})}_{\bm{3},3}}&{\beta Y^{(k_{2})}_{\bm{3},2}}&{\beta Y^{(k_{2})}_{\bm{3},1}}\\
	{\gamma Y^{(k_{3})}_{\bm{3},3}}&{\gamma Y^{(k_{3}}_{\bm{3},2}}&{\gamma Y^{(k_{3})}_{\bm{3},1}}\end{pmatrix}v$    \\ \cline{1-2} \cline{4-4} 

 $C^{(k_{1},k_{2},k_{3})}_{3}$ & $(\bm{1^{\prime\prime}},\bm{1^{\prime\prime}},\bm{1^{\prime\prime}})$ &  & $\begin{pmatrix}{\alpha Y^{(k_{1})}_{\bm{3},2}}&{\alpha Y^{(k_{1})}_{\bm{3},1}}&{\alpha Y^{(k_{1})}_{\bm{3},3}}\\
	{\beta Y^{(k_{2})}_{\bm{3},2}}&{\beta Y^{(k_{2})}_{\bm{3},1}}&{\beta Y^{(k_{2})}_{\bm{3},3}}\\
	{\gamma Y^{(k_{3})}_{\bm{3},2}}&{\gamma Y^{(k_{3})}_{\bm{3},1}}&{\gamma Y^{(k_{3})}_{\bm{3},3}}\end{pmatrix}v$    \\ \hline 

$C^{(k_{1},k_{2},k_{3})}_{4}$ & $(\bm{1},\bm{1},\bm{1^{\prime}})$ &   & $\begin{pmatrix}{\alpha Y^{(k_{1})}_{\bm{3},1}}&{\alpha Y^{(k_{1})}_{\bm{3},3}}&{\alpha Y^{(k_{1})}_{\bm{3},2}}\\
	{\beta Y^{(k_{2})}_{\bm{3},1}}&{\beta Y^{(k_{2})}_{\bm{3},3}}&{\beta Y^{(k_{2})}_{\bm{3},2}}\\
	{\gamma Y^{(k_{3})}_{\bm{3},3}}&{\gamma Y^{(k_{3})}_{\bm{3},2}}&{\gamma Y^{(k_{3})}_{\bm{3},1}}\end{pmatrix}v$    \\ \cline{1-2} \cline{4-4} 

$C^{(k_{1},k_{2},k_{3})}_{5}$ & $(\bm{1},\bm{1},\bm{1^{\prime\prime}})$ &   & $\begin{pmatrix}{\alpha Y^{(k_{1})}_{\bm{3},1}}&{\alpha Y^{(k_{1})}_{\bm{3},3}}&{\alpha Y^{(k_{1})}_{\bm{3},2}}\\
	{\beta Y^{(k_{2})}_{\bm{3},1}}&{\beta Y^{(k_{2})}_{\bm{3},3}}&{\beta Y^{(k_{2})}_{\bm{3},2}}\\
	{\gamma Y^{(k_{3})}_{\bm{3},2}}&{\gamma Y^{(k_{3})}_{\bm{3},1}}&{\gamma Y^{(k_{3})}_{\bm{3},3}}\end{pmatrix}v$    \\ \cline{1-2} \cline{4-4} 

$C^{(k_{1},k_{2},k_{3})}_{6}$ & $(\bm{1^{\prime}},\bm{1^{\prime}},\bm{1})$ & \multirow{4}{*}{$\begin{array}{cc}k_{{1}}<k_{{2}}\in\{\pm4,\pm2,0\},\\k_{{3}}\in\{\pm4,\pm2,0\}\end{array}$}  & $\begin{pmatrix}{\alpha Y^{(k_{1})}_{\bm{3},3}}&{\alpha Y^{(k_{1})}_{\bm{3},2}}&{\alpha Y^{(k_{1})}_{\bm{3},1}}\\
	{\beta Y^{(k_{2})}_{\bm{3},3}}&{\beta Y^{(k_{2})}_{\bm{3},2}}&{\beta Y^{(k_{2})}_{\bm{3},1}}\\
	{\gamma Y^{(k_{3})}_{\bm{3},1}}&{\gamma Y^{(k_{3})}_{\bm{3},3}}&{\gamma Y^{(k_{3})}_{\bm{3},2}}\end{pmatrix}v$    \\ \cline{1-2} \cline{4-4} 

$C^{(k_{1},k_{2},k_{3})}_{7}$ & $(\bm{1^{\prime}},\bm{1^{\prime}},\bm{1^{\prime\prime}})$ &    & $\begin{pmatrix}{\alpha Y^{(k_{1})}_{\bm{3},3}}&{\alpha Y^{(k_{1})}_{\bm{3},2}}&{\alpha Y^{(k_{1})}_{\bm{3},1}}\\
	{\beta Y^{(k_{2})}_{\bm{3},3}}&{\beta Y^{(k_{2})}_{\bm{3},2}}&{\beta Y^{(k_{2})}_{\bm{3},1}}\\
	{\gamma Y^{(k_{3})}_{\bm{3},2}}&{\gamma Y^{(k_{3})}_{\bm{3},1}}&{\gamma Y^{(k_{3})}_{\bm{3},3}}\end{pmatrix}v$    \\ \cline{1-2} \cline{4-4} 

$C^{(k_{1},k_{2},k_{3})}_{8}$ & $(\bm{1^{\prime\prime}},\bm{1^{\prime\prime}},\bm{1})$ &   & $\begin{pmatrix}{\alpha Y^{(k_{1})}_{\bm{3},2}}&{\alpha Y^{(k_{1})}_{\bm{3},1}}&{\alpha Y^{(k_{1})}_{\bm{3},3}}\\
	{\beta Y^{(k_{2})}_{\bm{3},2}}&{\beta Y^{(k_{2})}_{\bm{3},1}}&{\beta Y^{(k_{2})}_{\bm{3},3}}\\
	{\gamma Y^{(k_{3})}_{\bm{3},1}}&{\gamma Y^{(k_{3})}_{\bm{3},3}}&{\gamma Y^{(k_{3})}_{\bm{3},2}}\end{pmatrix}v$    \\ \cline{1-2} \cline{4-4} 

$C^{(k_{1},k_{2},k_{3})}_{9}$ & $(\bm{1^{\prime\prime}},\bm{1^{\prime\prime}},\bm{1^{\prime}})$ &   & $\begin{pmatrix}{\alpha Y^{(k_{1})}_{\bm{3},2}}&{\alpha Y^{(k_{1})}_{\bm{3},1}}&{\alpha Y^{(k_{1})}_{\bm{3},3}}\\
	{\beta Y^{(k_{2})}_{\bm{3},2}}&{\beta Y^{(k_{2})}_{\bm{3},1}}&{\beta Y^{(k_{2})}_{\bm{3},3}}\\
	{\gamma Y^{(k_{3})}_{\bm{3},3}}&{\gamma Y^{(k_{3})}_{\bm{3},2}}&{\gamma Y^{(k_{3})}_{\bm{3},1}}\end{pmatrix}v$    \\ \hline

$C^{(k_{1},k_{2},k_{3})}_{10}$ & $(\bm{1},\bm{1^{\prime}},\bm{1^{\prime\prime}})$ & $k_{{1}},k_{{2}},k_{{3}}\in\{\pm4,\pm2,0\}$  & $\begin{pmatrix}{\alpha Y^{(k_{1})}_{\bm{3},1}}&{\alpha Y^{(k_{1})}_{\bm{3},3}}&{\alpha Y^{(k_{1})}_{\bm{3},2}}\\
	{\beta Y^{(k_{2})}_{\bm{3},3}}&{\beta Y^{(k_{2})}_{\bm{3},2}}&{\beta Y^{(k_{2})}_{\bm{3},1}}\\
	{\gamma Y^{(k_{3})}_{\bm{3},2}}&{\gamma Y^{(k_{3})}_{\bm{3},1}}&{\gamma Y^{(k_{3})}_{\bm{3},3}}\end{pmatrix}v$    \\ \hline
 \end{tabular}
\caption{\label{tab:sum_ch}Possible assignments for the $A_{4}$ representations and modular weights of the lepton fields. The LH doublets $L$ form a triplet $\bm{3}$ under $A_4$ with modular weight $k_{L}$. The RH charged leptons $E^{c}_{i}$ ($i=1,2,3$) transform under $A_{4}$ as $\rho_{E^{c}_{i}}$ with modular weights $k_{E^{c}_{i}}$. The polyharmonic Maa{\ss}  form triplet $Y^{(k_{i})}_{\bm{3}}$ satisfies $k_{i} = k_{L} + k_{E^{c}_{i}}$, and $v = \langle H \rangle$ denotes the vacuum expectation value of the Higgs field.
}
\end{center}
\end{table}

\subsection{\label{sec:nu_mass}Neutrino sector}

In the present work, the neutrinos are assumed to be Majorana particles, and their masses are described by the effective Weinberg operator. 
According to the symmetry constraints of the model, when the LH lepton doublets $L$ transform as the triplet $\bm{3}$ under $A_4$, the antisymmetric combination $(LL)_{\bm{3_A}}$ vanishes. Therefore, nonzero neutrino masses require either a singlet polyharmonic Maa{\ss} form $Y^{(2k_L)}_{\bm{1}}$, $Y^{(2k_L)}_{\bm{1^{\prime}}}$, $Y^{(2k_L)}_{\bm{1^{\prime\prime}}}$, or a triplet polyharmonic Maa{\ss} form $Y^{(2k_L)}_{\bm{3}}$. Then the most general modular invariant Lagrangian for neutrino masses can be written as
\begin{equation}\label{eq:weinberg}
\hskip-0.08in  W_{j}~:~-\mathcal{L}_{\nu}=	\frac{H^{2}}{2\Lambda}\left[g_{1}Y^{(k_{j})}_{\bm{1}}\left(LL\right)_{\bm{1}}+g_{2}\left(\left(LL\right)_{\bm{3_{S}}}Y^{(k_{j})}_{\bm{3}}\right)_{\bm{1}}+g_{3}Y^{(k_{j})}_{\bm{1^{\prime}}}\left(LL\right)_{\bm{1^{\prime\prime}}}
+g_{4}Y^{(k_{j})}_{\bm{1^{\prime\prime}}}\left(LL\right)_{\bm{1^{\prime}}}\right]\,.
\end{equation}
where we consider, for simplicity, four distinct cases $W_j$ ($j= 1, 2, 3, 4$) corresponding to the modular weights  $k_{j}=2k_{L}=-4,-2,0,2$, respectively. In these four cases, the last two terms in Eq.~\eqref{eq:weinberg} vanish because there are no corresponding polyharmonic Maa{\ss} forms $Y^{(k_{j})}_{\bm{1^{\prime}}}$ and $Y^{(k_{j})}_{\bm{1^{\prime\prime}}}$ at those weights. Consequently, the resulting light neutrino mass matrix depends only on the two parameters $g_1$ and $g_2$. By applying the decomposition rule of the finite modular group $A_{4}$, the light neutrino mass matrix can be written in the following form:
\begin{equation}\label{eq:nu_mass}
	M_{\nu}(k_{j})=\frac{{v}^2}{2\Lambda}\begin{pmatrix}{g_{1}Y^{(k_{j})}_{\bm{1}}+{2g_{2}}Y^{(k_{j})}_{\bm{3},1}}&{{-g_{2}}Y^{(k_{j})}_{\bm{3},3}}&{{-g_{2}}Y^{(k_{j})}_{\bm{3},2}}\\
	{{-g_{2}}Y^{(k_{j})}_{\bm{3},3}}&{{2g_{2}}Y^{(k_{j})}_{\bm{3},2}}&{g_{1}Y^{(k_{j})}_{\bm{1}}-{g_{2}}Y^{(k_{j})}_{\bm{3},1}}\\
	{{-g_{2}}Y^{(k_{j})}_{\bm{3},2}}&{g_{1}Y^{(k_{j})}_{\bm{1}}-{g_{2}}Y^{(k_{j})}_{\bm{3},1}}&{{2g_{2}}Y^{(k_{j})}_{\bm{3},3}}\end{pmatrix}, 
\end{equation}
where the phase of $g_{1}$ can be eliminated by field redefinition, while the phase of $g_{2}$  remains physical and cannot be removed. When gCP symmetry is imposed, both  $g_{1}$ and $g_{2}$ are real in our working basis.
	
\subsection{Numerical results}

\begin{table}[t!]
\centering
\begin{tabular}{|c|c|c||c|c|}\hline \hline
\multirow{2}{*}{Observables}  &  	\multicolumn{2}{c||}{NO}   &      \multicolumn{2}{c|}{IO}       \\ \cline{2-3} \cline{4-5}
	
& $\text{bf}\pm1\sigma$  & $3\sigma$ region & $\text{bf}\pm1\sigma$ & $3\sigma$ region   \\ \hline

&   &  &  &    \\[-0.150in]
	
$\sin^2\theta_{13}$ & $0.02215^{+0.00056}_{-0.00058}$ & $[0.02030,0.02388]$ & $0.02231^{+0.00056}_{-0.00056}$ &  $[0.02060,0.02409]$  \\ [0.050in]
	
$\sin^2\theta_{12}$ & $0.308^{+0.012}_{-0.011}$ & $[0.275,0.345]$ & $0.308^{+0.012}_{-0.011}$ & $[0.275,0.345]$  \\ [0.050in]
	
$\sin^2\theta_{23}$  & $0.470^{+0.017}_{-0.013}$  & $[0.435,0.585]$ & $0.550^{+0.012}_{-0.015}$  & $[0.440,0.584]$   \\ [0.050in]
	
$\delta_{CP}/\pi$  & $1.178^{+0.144}_{-0.228}$ & $[0.689,2.022]$  & $1.522^{+0.122}_{-0.139}$ & $[1.117,1.861]$  \\ [0.050in]

$\frac{\Delta m^2_{21}}{10^{-5}\text{eV}^2}$ & $7.49^{+0.19}_{-0.19}$ & $[6.92,8.05]$ & $7.49^{+0.19}_{-0.19}$ & $[6.92,8.05]$  \\ [0.050in]

$\frac{\Delta m^2_{3\ell}}{10^{-3}\text{eV}^2}$ & $2.513^{+0.021}_{-0.019}$ & $[2.451,2.578]$ & $-2.484^{+0.020}_{-0.020}$ & $[-2.547,-2.421]$ \\ [0.050in]

$\Delta m^2_{21}/\Delta m^2_{3\ell}$  &  $0.0298^{+0.00079}_{-0.00079}$  & $[0.0268,0.0328]$ & $-0.0302^{+0.00080}_{-0.00080}$ & $[-0.0333,-0.0272]$  \\ [0.050in]

$m_e/m_{\mu}$  & $0.004737$ & --- & $0.004737$ & --- \\ [0.050in]

$m_{\mu}/m_{\tau}$  & $0.05882$ & --- & $0.05882$ & --- \\ [0.050in]
 $m_{e}/\text{MeV}$ & $0.469652$  &  --- & $0.469652$  &  ---  \\ [0.050in] \hline \hline
	
\end{tabular}
\caption{\label{tab:bf_13sigma_data}
The best fit values, $1\sigma$ and $3\sigma$ ranges for the mixing parameters and lepton mass ratios are presented, where the experimental data and uncertainties for both the NO and IO neutrino mass spectra are sourced from NuFIT 6.0 with Super-Kamiokande atmospheric data~\cite{Esteban:2024eli}. It is important to note that $\Delta m^2_{3\ell}=\Delta m^2_{31}>0$ for NO and $\Delta m^2_{3\ell}=\Delta m^2_{32}<0$ for IO. The $1\sigma$ uncertainties for the charged lepton mass ratios are considered to be $0.1\%$ of their central values in the $\chi^2$ analysis.}
\end{table}

In sections~\ref{sec:ch_mass} and~\ref{sec:nu_mass}, we have separately discussed the representation and weight assignments, along with the resulting mass matrices for the charged leptons and neutrinos, respectively. We assume that the LH doublet leptons transform as the $A_{4}$ triplet $\bm{3}$, while the RH charged leptons are assigned to $A_{4}$ singlets. Considering the ten possible representation assignments for the RH charged lepton fields and polyharmonic Maa{\ss} form multiplets $Y^{(k)}_{\bm{r}}$ within modular weights $-4 \leq k \leq 4$, we find that 455 specific assignments lead to charged lepton mass matrices describable by only three real parameters. The corresponding representation and weight assignments, along with all such mass matrices, are summarized in table~\ref{tab:sum_ch}. In the charged lepton sector, $k_{L}$ is relatively unconstrained, whereas the sum $k_{i}=k_{E^c_i} + k_{L}$ is fixed as shown in the table. Neutrino masses are generated via the effective Weinberg operator. Corresponding to the weight $k_{L}$, only four distinct assignments yield light neutrino mass matrices that are described by the minimal set of free parameters. The corresponding mass matrices are given in Eq.~\eqref{eq:nu_mass}. 

When building explicit lepton models, the representation and modular weight assignments of the lepton doublet $L$ must be chosen consistently in both the charged lepton and neutrino sectors. As noted above, we take the left-handed leptons to transform as the $A_{4}$ triplet $\bm{3}$ in both sectors.
By combining the possible structures from the charged lepton mass matrices in table~\ref{tab:sum_ch} and the neutrino mass matrices in Eq.~\eqref{eq:nu_mass}, we obtain a total of $455 \times 4 = 1820$ minimal non-supersymmetric lepton models based on the finite modular group $A_{4}$. These models are labeled as $\{C^{(k_1,k_2,k_3)}_{m}, W_{j}\}$, where $m = 1, 2, \dots, 10$ and $j = 1, 2, 3, 4$.
For each model, the modular weights of the matter fields are uniquely determined, though their explicit values are not listed here. Note that the coupling constants $\alpha$, $\beta$ and $\gamma$ in the charged lepton mass matrices can be chosen as positive real numbers. The light neutrino mass matrices contain two additional real parameters $\left|g_{2}/g_{1}\right|$ and $\text{arg}(g_{2}/g_{1})$ besides the overall factor $g_{1}v^2/\Lambda$ and the modulus $\tau$. Thus, in the absence of gCP symmetry, all 1820 models involve eight independent real parameters. When gCP symmetry compatible with $A_4$ is imposed, the ratio $g_2/g_1$ is further constrained to be real, reducing the number of real input parameters to seven.

For each lepton flavor model, we must verify whether it can reproduce the experimental data within uncertainties. To this end, we perform a systematic numerical and $\chi^2$ analysis of all 1820 minimal lepton models with and without gCP symmetry, and for both NO and IO neutrino mass spectra. The $\chi^2$ function is defined in the standard form:
\begin{equation}\label{eq:chisq}
    \chi^2=\sum_{i=1}^7\left(\frac{P_{i}(x)-O_{i}}{\sigma_{i}}\right)^2,
\end{equation}
where $O_i$ and $\sigma_i$ denote the central values and $1\sigma$ uncertainties of the seven  dimensionless observables:
\begin{equation}\label{eq:six_output}
m_e/m_\mu, \quad m_\mu/m_\tau, \quad \sin^2\theta_{12}, \quad \sin^2\theta_{13}, \quad \sin^2\theta_{23}, \quad \delta_{CP}, \quad \Delta m_{21}^2 / \Delta m_{3l}^2,
\end{equation}
as listed in table~\ref{tab:bf_13sigma_data}, where the lepton mixing parameters are taken from the NuFIT 6.0 with inclusion of the data on atmospheric neutrinos provided by the Super-Kamiokande~\cite{Esteban:2024eli}. The $P_i(x)$ in Eq.~\eqref{eq:chisq} represent the model predictions for these observables given a set of input parameters $x$. When gCP symmetry is not imposed, $x$ consists of the six parameters:
\begin{equation}\label{eq:six_inputs}
\Re\tau, \quad \Im\tau, \quad \beta/\alpha, \quad \gamma/\alpha, \quad |g_2/g_1|, \quad \arg(g_2/g_1).
\end{equation}
If gCP symmetry is included, $\arg(g_2/g_1)$ is restricted to $0$ or $\pi$, reducing the number of free parameters to five. In this work, the absolute values of the Yukawa coupling ratios are uniformly sampled in $[0,10^6]$, and the phase $\arg(g_2/g_1)$ in $[0,2\pi)$.  The complex modulus $\tau$ is restricted to lie in the fundamental domain $\mathcal{D}=\left\{\tau\in\mathcal{H}\Big| -\frac{1}{2}\leq\Re(\tau)\leq\frac{1}{2},~|\tau|\geq1\right\}$, since the underlying theory has the modular symmetry $\overline{\Gamma}$, and consequently vacua related by modular transformations are physically equivalent~\cite{Novichkov:2018ovf}. It is particularly noteworthy that the overall parameter $\alpha v$ of the charged lepton mass matrix and the normalization factor $g_{1}v^{2} /\Lambda$ of the light neutrino mass matrix can be determined by the experimentally measured electron mass $m_e$ and the solar neutrino mass squared difference $\Delta m_{21}^2$, respectively.

For each set of input parameters, we can calculate the corresponding predictions for lepton masses, mixing parameters, $\chi^2$ function, as well as the effective mass $m_{\beta\beta}$ in neutrinoless double beta decay ($0\nu\beta\beta$-decay) and the kinematical mass $m_{\beta}$ in beta decay which are defined as:
\begin{equation}
m_{\beta\beta}=\left|\sum^3_{i}m_{i}U^{2}_{1i}\right|\,, \qquad m_{\beta}=\left[\sum^3_{i}m^2_{i}\left|U_{1i}\right|^{2}\right]^{1/2}\,,
\end{equation}
where $m_i$ are the light neutrino masses and $U$ is the PMNS matrix. We adopt the standard parametrization of the PMNS matrix~\cite{ParticleDataGroup:2024cfk}:
\begin{equation}\label{eq:PMNS}
U=\begin{pmatrix}
c_{12}c_{13}  &   s_{12}c_{13}   &   s_{13}e^{-i\delta_{CP}}  \\
-s_{12}c_{23}-c_{12}s_{13}s_{23}e^{i\delta_{CP}}   &  c_{12}c_{23}-s_{12}s_{13}s_{23}e^{i\delta_{CP}}  &  c_{13}s_{23}  \\
s_{12}s_{23}-c_{12}s_{13}c_{23}e^{i\delta_{CP}}   & -c_{12}s_{23}-s_{12}s_{13}c_{23}e^{i\delta_{CP}}  &  c_{13}c_{23}
\end{pmatrix}\text{diag}(1,e^{i\frac{\alpha_{21}}{2}},e^{i\frac{\alpha_{31}}{2}})\,,
\end{equation}
where $c_{ij}\equiv \cos\theta_{ij}$, $s_{ij}\equiv \sin\theta_{ij}$, $\delta_{CP}$ is the Dirac CP-violating phase, and $\alpha_{21,31}$ are Majorana CP-violating phases. 

\begin{figure}[t!]
	\centering
	\begin{tabular}{c}
		\includegraphics[width=0.495\linewidth]{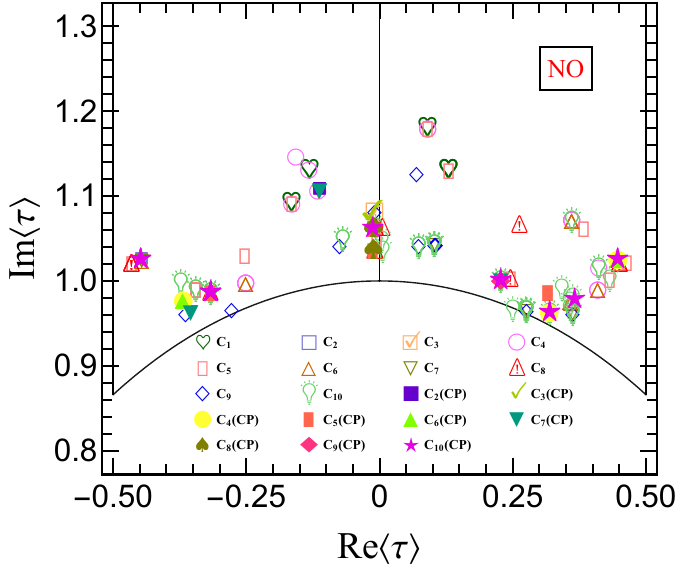} 
\includegraphics[width=0.495\linewidth]{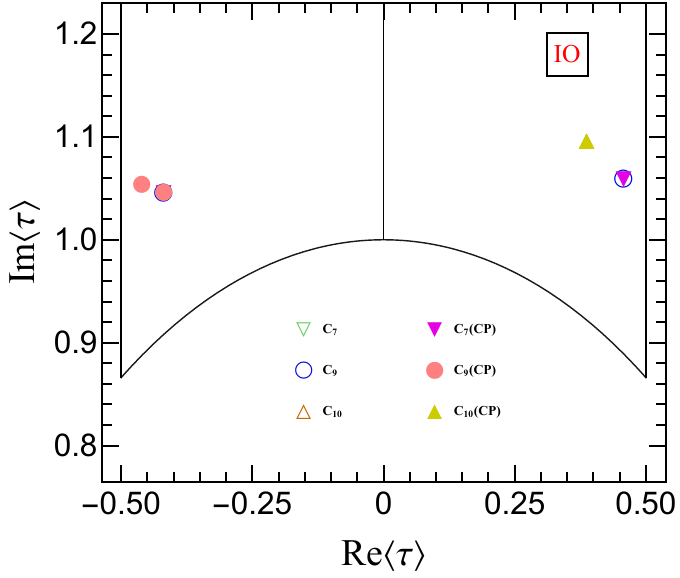}
	\end{tabular}
\textcolor{red}{	\caption{\label{fig:bf_tau}  The best fit values of modulus $\tau$ for 147 (47)   and 6 (5) viable models in the case without (with) gCP symmetry for NO and IO neutrino mass spectra, respectively. }}
	\end{figure}

We will now compare the predictions of the 1820 models with the latest NuFIT results. A model is phenomenologically viable if its predicted lepton mass ratios and mixing parameters at the $\chi^{2}$ minimum fall within the $3\sigma$ intervals provided in table~\ref{tab:bf_13sigma_data}. Furthermore, our analysis selects models whose predictions must also satisfy the constraint that the total neutrino mass $\sum_{i=1}^{3} m_{i}$ must be below the Planck $+$ lensing $+$ BAO~ limit of 120 meV~\cite{Planck:2018vyg}. Through comprehensive scanning of the input parameter space, we systematically determine the minimal $\chi^2$ values.  From our analysis of 1820 models without gCP symmetry, we find that 147 are compatible with experimental data in the NO mass spectrum, while only 6 are compatible in the IO. The complete fitting results for these viable models, including input parameters, mixing angles, CP-violating phases, neutrino masses, the effective mass $m_{\beta\beta}$ in $0\nu\beta\beta$-decay and the kinematical mass $m_{\beta}$  in beta decay are summarized in  table~\ref{tab:WO_NO_bf_noGCP} and table~\ref{tab:WO_IO_bf_noGCP}.  Only 47 out of 147 models for NO and 5 out of 6 for IO produce results compatible with experimental data on lepton mixing parameters and masses, when the finite modular group $A_{4}$ is extended to include gCP symmetry. The best fit values of input parameters, mixing parameters and lepton masses  are presented in table~\ref{tab:WO_NO_bf_GCP} and table~\ref{tab:WO_IO_bf_GCP}. We find that five models without gCP symmetry and in NO case—namely $\{C^{(-2,0,2)}_{2}, W_{1}\}$, $\{C^{(-2,0,4)}_{2}, W_{1}\}$, $\{C^{(-2,0,-2)}_{6}, W_{1}\}$, $\{C^{(-2,0,0)}_{6}, W_{1}\}$ and $\{C^{(-4,4,-4)}_{7}, W_{1}\}$ are disfavored once the current JUNO 59.1‑day constraint on $\sin^{2}\theta_{12}$ is applied~\cite{JUNO:2025gmd}. Their predicted values lie within the $3\sigma$ interval
\begin{equation}\label{eq:s12sq_3sigma_JUNO}
0.2831 \leq \sin^{2}\theta_{12} \leq 0.3353\,,
\end{equation}
which is derived from the central value $\sin^2\theta_{12}=0.3092\pm0.0087$ by subtracting three times the $1\sigma$ uncertainty. All remaining models remain compatible with current oscillation data, including this JUNO measurement. 
Note that the Weinberg operator model discussed in Ref.~\cite{Qu:2024rns} corresponds to our model $\{C_{10}^{(-2,0,0)}, W_{1}\}$. This model does not appear in table~\ref{tab:WO_IO_bf_noGCP} for the IO case because it fails to satisfy the cosmological upper bound on the sum of neutrino masses $\sum_{i=1}^{3} m_{i} < 120~\text{meV}$  which is enforced in our analysis. 

\begin{figure}[t!]
	\centering
	\begin{tabular}{c}
		\includegraphics[width=0.925\linewidth]{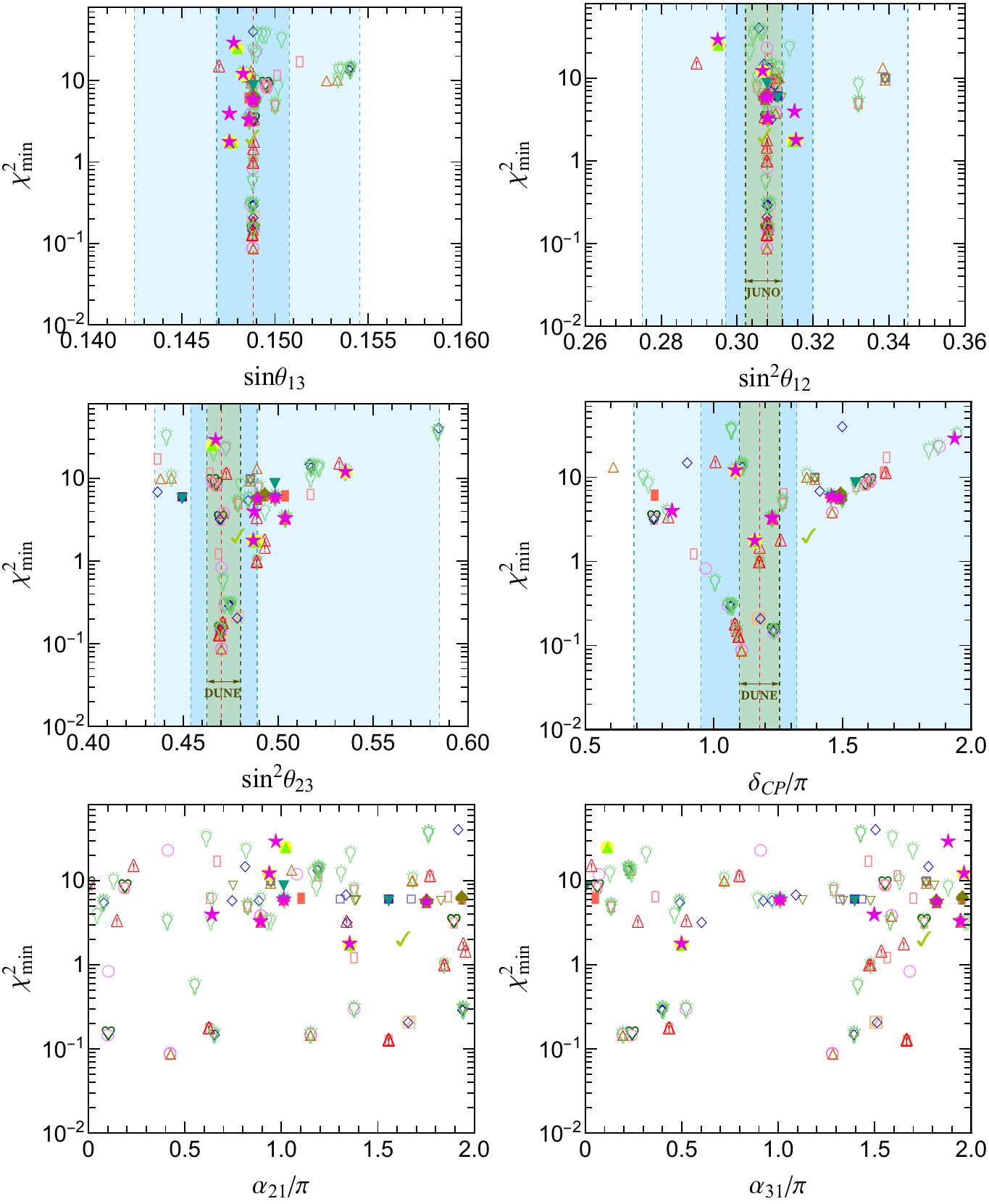}
	\end{tabular}
	\caption{\label{fig:bf_mixing_NO}  {The results of the best fit values of the  minimum value of $\chi^2$, the three lepton mixing angles and three CP-violating phases for viable models without (147 models) and with (47 models) gCP symmetry in the NO case. The red dashed lines in the first four panels represent the best fit values, and the light blue bounds represent the $1\sigma$ and $3\sigma$ ranges from NuFIT 6.0 with Super-Kamiokande atmospheric data~\cite{Esteban:2020cvm}. The lighter green band in the panel of $\sin^{2}\theta_{12}$ is the prospective $3\sigma$ range after 6 years of JUNO running~\cite{JUNO:2022mxj}. The lighter green regions in the panels of $\sin^{2}\theta_{23}$ and $\delta_{CP}$ are the resolution in degrees after 15 years of DUNE running~\cite{DUNE:2020ypp} for true values of them corresponding to their  best fit values  given by NuFIT 6.0.} }
\end{figure}

\begin{figure}[t!]
	\centering
	\begin{tabular}{c}
		\includegraphics[width=0.98\linewidth]{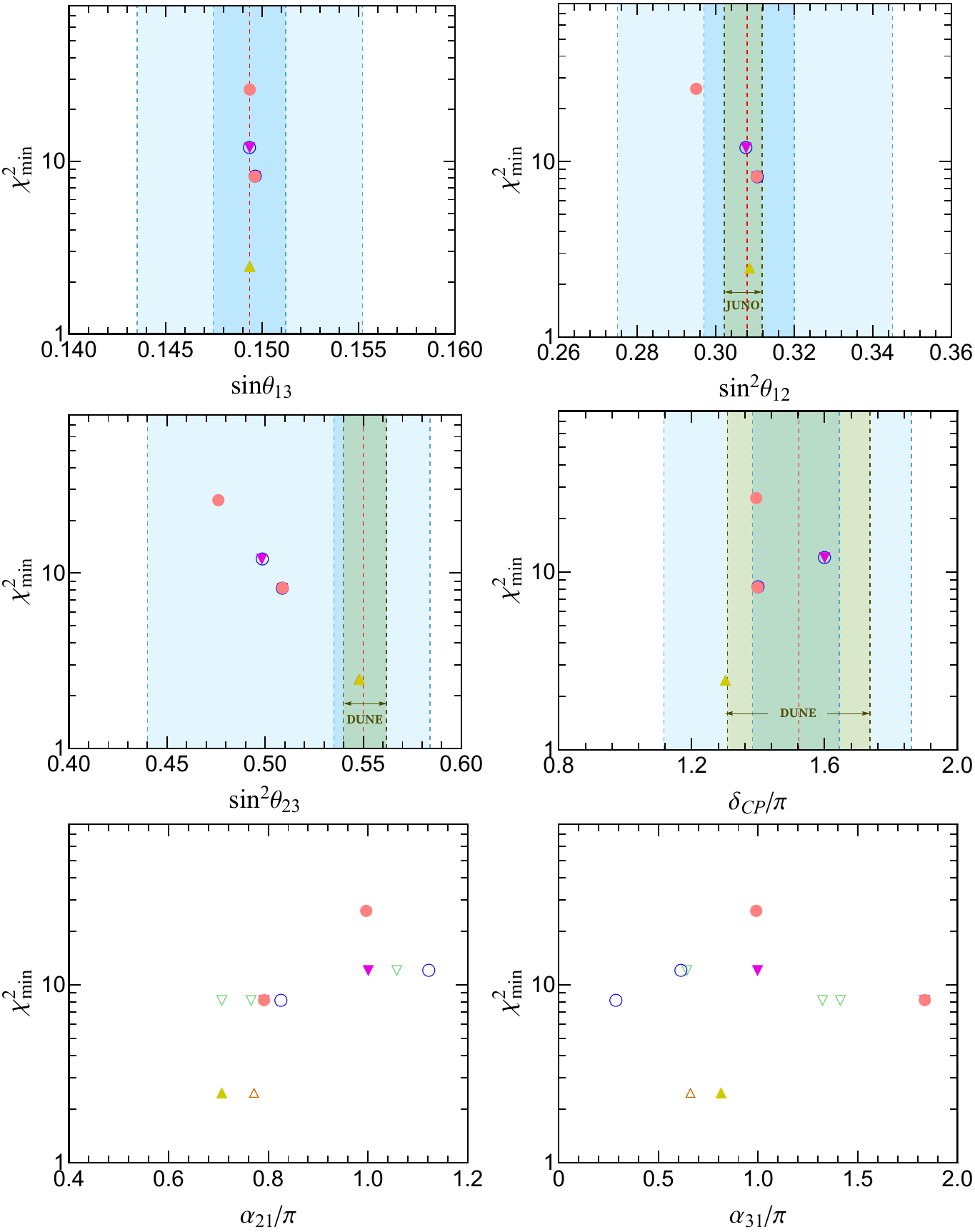}
	\end{tabular}
	\caption{\label{fig:bf_mixing_IO}  {The results of the best fit values of the  minimum value of $\chi^2$, the three lepton mixing angles and three CP-violating phases for all viable models without and with gCP symmetry in the IO case. } }
\end{figure}

\begin{figure}[t!]
	\centering
	\begin{tabular}{c}
		\includegraphics[width=0.99\linewidth]{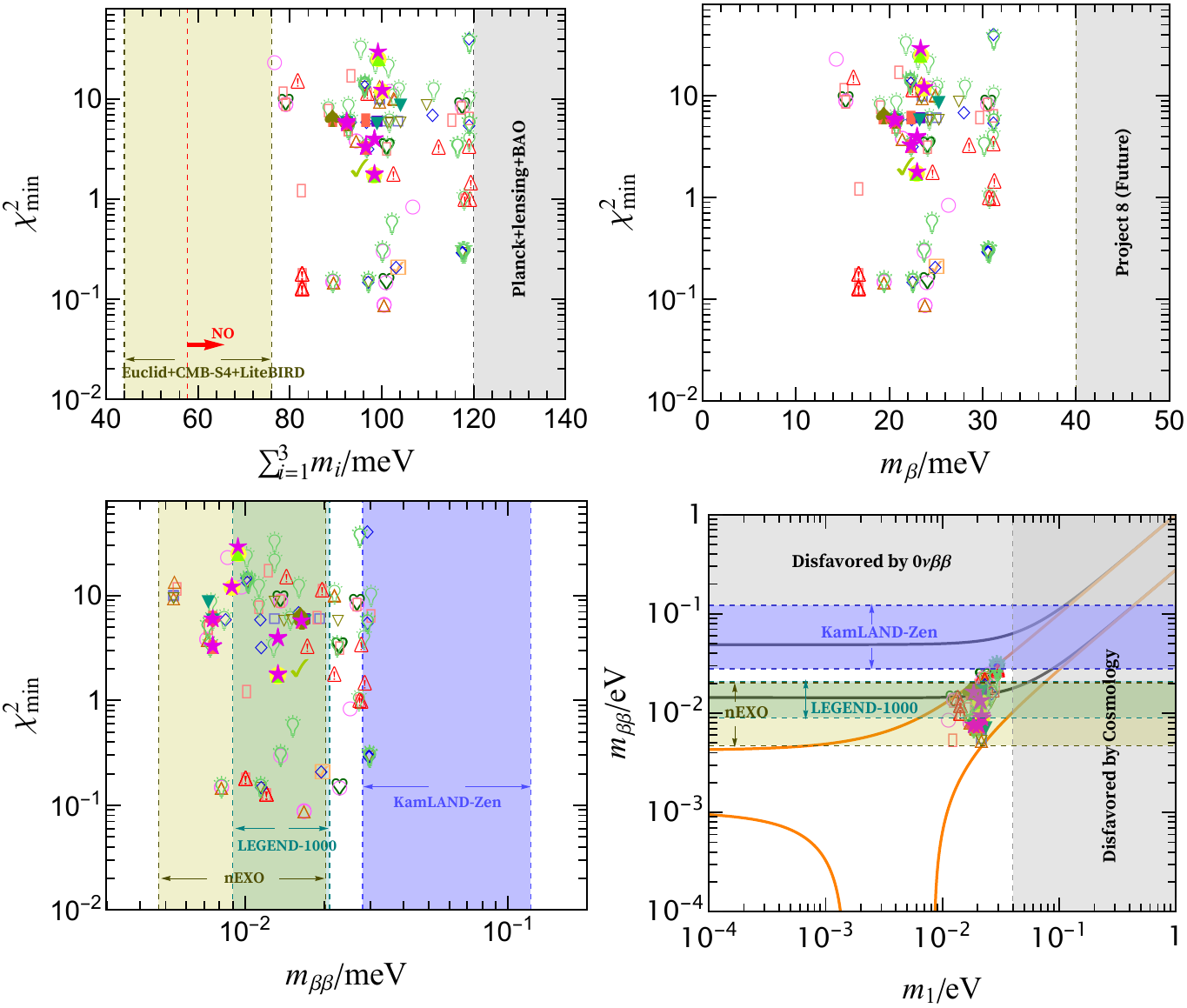}
	\end{tabular}
	\caption{\label{fig:bf_mass_NO}  The best fit values of the  minimum value of $\chi^2$, the effective mass $m_{\beta\beta}$ in $0\nu\beta\beta$-decay, the kinematical mass $m_{\beta}$ in beta decay  and the three neutrino mass sum $\sum_{i=1}^{3} m_{i}$. In the panel of the neutrino mass sum $\sum_{i=1}^{3} m_{i}$, the vertical bands indicate the current most stringent limit $\sum_{i=1}^{3} m_{i}<120\,\text{meV}$ from the Planck $+$ lensing $+$ BAO~\cite{Planck:2018vyg},  the next-generation experiments sensitivity ranges $\sum_{i=1}^{3} m_{i}<(44-76)\,\text{meV}$ of Euclid+CMB-S4+LiteBIRD~\cite{Euclid:2024imf}, and the red dashed line represents the limitation of the NO case ($\sum_{i=1}^{3} m_{i}\geq 57.75\,\text{meV}$). In the panel of the kinematical mass $m_{\beta}$ in beta decay, the gray region represents Project 8 future bound ($m_{\beta}<0.04\,\text{meV}$)~\cite{Project8:2022wqh}. In the panel of the effective Majorana mass $m_{\beta\beta}$, the vertical bands indicate the latest result $m_{\beta\beta}<(28-122)\,\text{meV}$ of KamLAND-Zen~\cite{KamLAND-Zen:2024eml}, and the next-generation experiments sensitivity ranges $m_{\beta\beta}<(9-21)\,\text{meV}$ from LEGEND-1000~\cite{LEGEND:2021bnm} and $m_{\beta\beta}<(4.7-20.3)\,\text{meV}$ from nEXO~\cite{nEXO:2021ujk}.} 
\end{figure}

\begin{figure}[t!]
	\centering
	\begin{tabular}{c}
		\includegraphics[width=0.99\linewidth]{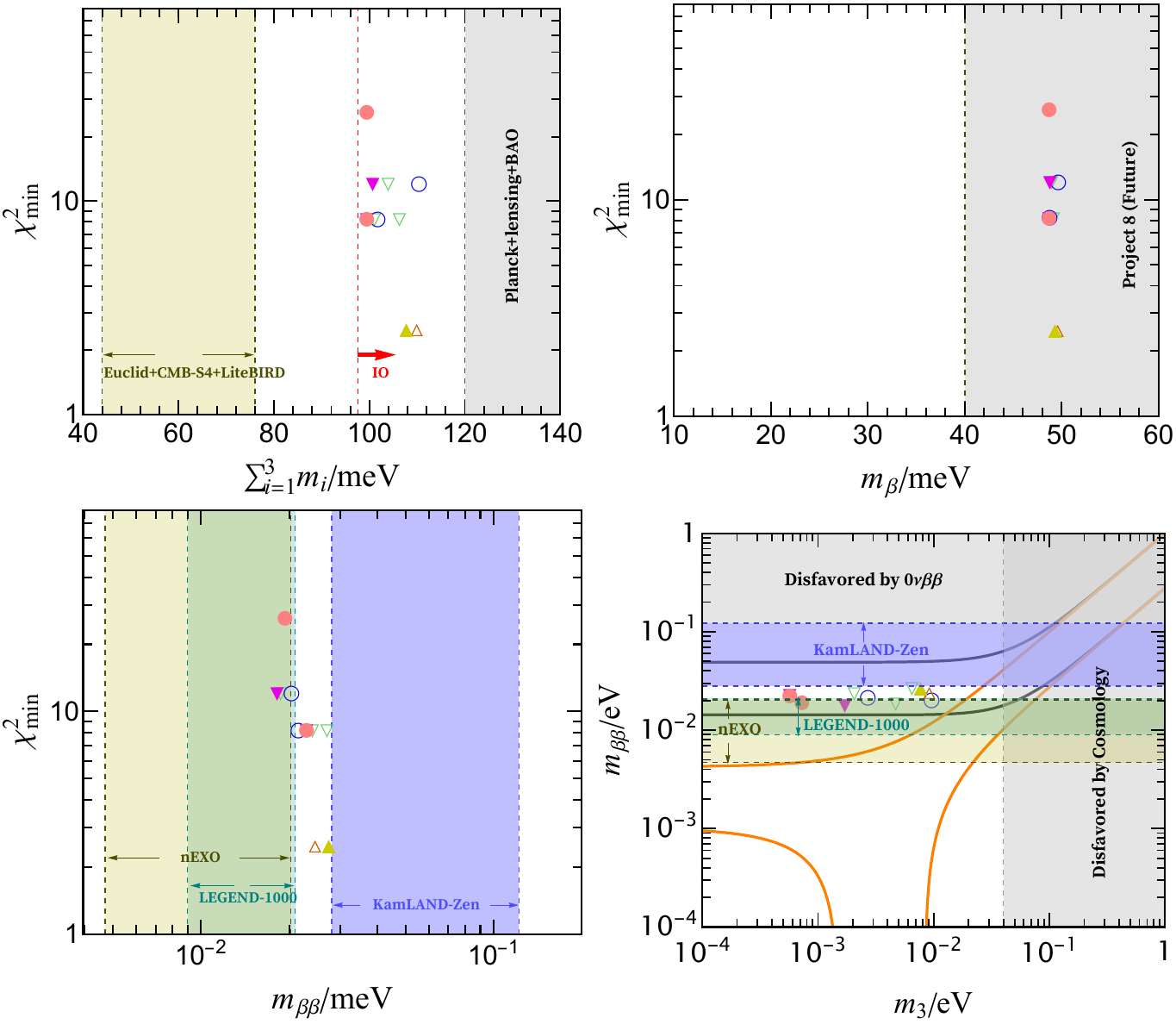}
	\end{tabular}
	\caption{\label{fig:bf_mass_IO}  The best fit values of the  minimum value of $\chi^2$, the effective mass $m_{\beta\beta}$, the kinematical mass $m_{\beta}$ and the mass sum $\sum_{i=1}^{3} m_{i}$. } 
\end{figure}

In all these models, the modulus $\tau$ is regarded as a free parameter to maximize the agreement between theoretical predictions and experimental data.  
The best fit values of the complex modulus $\tau$ in the fundamental domain of $\mathrm{SL}(2,\mathbb{Z})$ are displayed in figure~\ref{fig:bf_tau}. This analysis includes 194 viable models for the NO case (147 without CP and 47 with CP) and 11 for the IO case (6 without CP and 5 with CP). We find that the VEVs of $\tau$ in most viable models cluster near the regions $\Re\tau=0$, $|\tau|=1$ or $\Re \tau=\pm0.5$ for both mass orderings.
The key observables for each model, which include the three lepton mixing angles, three CP-violating phases, $\sum_{i=1}^{3} m_{i}$, $m_{\beta\beta}$ and $m_\beta$ are relevant  to their minimal $\chi^2$ values. The corresponding results for the NO and IO cases are displayed in figures~\ref{fig:bf_mixing_NO} and~\ref{fig:bf_mass_NO}, and figures~\ref{fig:bf_mixing_IO} and~\ref{fig:bf_mass_IO}, respectively. Only few models feature best fit values for all three mixing angles and the Dirac CP phase within the experimental $1\sigma$ ranges, indicating strong agreement with data, as shown in figures~\ref{fig:bf_mixing_NO} and~\ref{fig:bf_mixing_IO}.
The next-generation neutrino oscillation experiments and cosmological surveys will provide precise measurements of the lepton mixing parameters and neutrino masses. Combined with accurate determinations of $m_{\beta\beta}$, the joint analysis of these experimental data will offer crucial evidence for investigating various modular models.
Assuming the current best fit value for $\sin^{2}\theta_{12}$ remains unchanged, the next-generation JUNO experiment will measure this parameter with high precision after six years of data collection. As shown in figures~\ref{fig:bf_mixing_NO} and~\ref{fig:bf_mixing_IO}, its projected $3\sigma$ uncertainty range is wide enough to encompass the predicted $\sin^{2}\theta_{12}$ values of almost all models considered. Distinguishing among these modular models will require improved precision on $\theta_{23}$ and $\delta_{CP}$. Future long baseline experiments DUNE~\cite{DUNE:2020ypp} and T2HK~\cite{Hyper-Kamiokande:2018ofw} will critically test viable modular models through precision measurements of $\theta_{23}$ and $\delta_{CP}$. The projected angular resolution of these parameters in DUNE after 15 years running is also shown in the two figures. As these projections indicate, once DUNE and T2HK achieve their target sensitivity, the majority of the currently viable models will be decisively tested.  The combination of JUNO, DUNE and T2HK will provide a powerful approach to testing these models. Given the projected constraints from these experiments, only a small number of models remain consistent with data, while the vast majority will be disfavored.

Figures~\ref{fig:bf_mass_NO} and \ref{fig:bf_mass_IO} indicate that the predicted sum of neutrino masses $\sum_{i=1}^{3} m_{i}$ for all viable models lies within the detection range of future cosmological surveys. These estimates are consistent with the expected sensitivity of $\sum_{i=1}^{3} m_{i} < (44 - 76)\ \text{meV}$ from the combined Euclid+CMB-S4+LiteBIRD~\cite{Euclid:2024imf}. For the NO case, all viable models yield $m_{\beta}$ values below the forecasted detection limit of $0.04\ \text{eV}$ by Project 8~\cite{Project8:2022wqh}, while for the IO case, $m_{\beta}$ lies above this threshold. Predictions for $m_{\beta\beta}$ in all viable NO models are compatible with the current KamLAND-Zen constraint of $m_{\beta\beta} < (28 - 122)\ \text{meV}$~\cite{KamLAND-Zen:2024eml}, though this bound is expected to accommodate some viable IO models. Upcoming experiments such as LEGEND-1000 (aiming for $m_{\beta\beta} < (9 \sim 21)\ \text{meV}$)~\cite{LEGEND:2021bnm} and nEXO (with a projected reach of $m_{\beta\beta} < (4.7 \sim 20.3)\ \text{meV}$)~\cite{nEXO:2021ujk} will be sensitive enough to test most of the NO models and all of the IO models. The relationship between the lightest neutrino mass ($m_1$ or $m_3$) and $m_{\beta\beta}$ for each viable case is illustrated in figures~\ref{fig:bf_mass_NO} and~\ref{fig:bf_mass_IO}, highlighting parameter trends across mass orderings.

\section{\label{sec:example-models}Typical model}

\begin{figure}[t!]
	\centering
	\begin{tabular}{c}
		\includegraphics[width=0.99\linewidth]{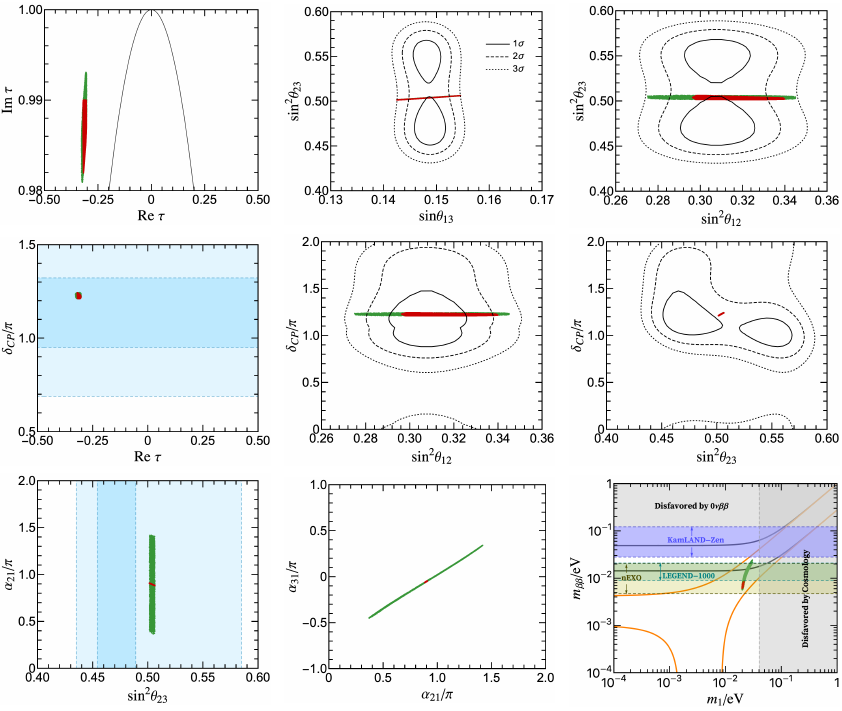}
	\end{tabular}
	\caption{\label{fig:WO_mixing_NO}  {The predicted correlations among the input parameters, lepton mixing angles and CP-violating phases in the model $\{C^{(2,4,2)}_{10},W_{3}\}$ with (red) and without (green) gCP symmetry for NO neutrino mass spectrum.} }
\end{figure}

\begin{figure}[t!]
	\centering
	\begin{tabular}{c}
		\includegraphics[width=0.99\linewidth]{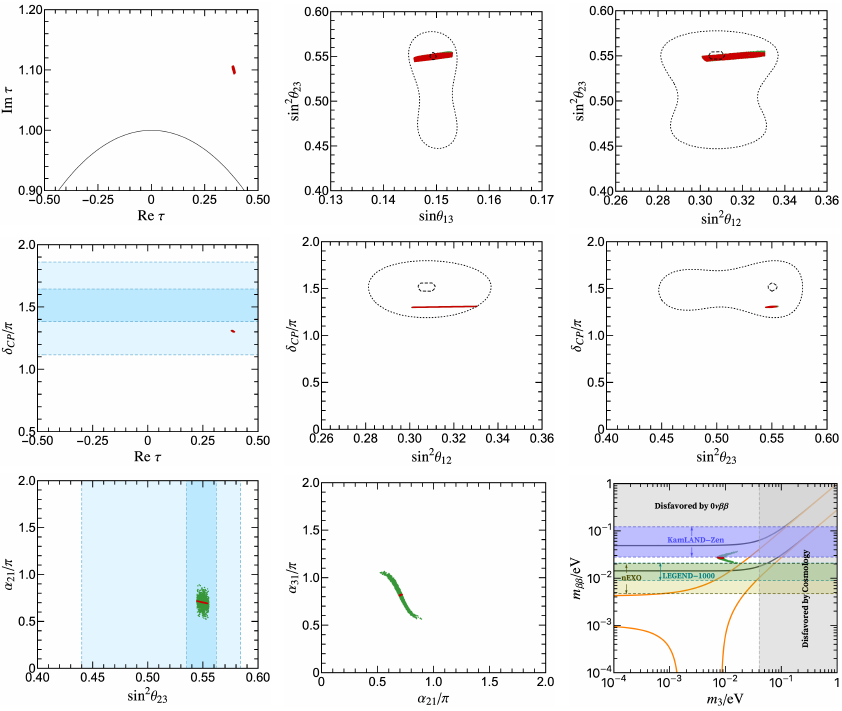}
	\end{tabular}
	\caption{\label{fig:WO_mixing_IO}  {The predicted correlations among the input parameters, lepton mixing angles and CP-violating phases in the model $\{C^{(2,4,2)}_{10},W_{3}\}$ with (red) and without (green) gCP symmetry for IO neutrino mass spectrum.} }
\end{figure}

Based on the preceding analysis, non-holomorphic $A_{4}$ modular models are highly predictive due to their limited free parameters, which naturally correlates lepton masses with mixing parameters.  This is particularly evident in models with gCP symmetry, where all six lepton masses and six mixing parameters are determined by just seven real input parameters. When gCP symmetry is absent, the number of input parameters increases to eight. To demonstrate the predictive capability of non-holomorphic $A_{4}$ modular invariant models, we present detailed numerical results for the representative model $\{C^{(2,4,2)}_{10},W_{3}\}$ as an illustrative example.
The representation and weight assignments for the lepton fields in this model are as follows:
\begin{equation}
L\sim3,\quad E_1^c\sim1,\quad E_2^c\sim1^{\prime},\quad E_3^c\sim1^{\prime\prime},\quad k_{L}=0, \quad k_{E^{c}_{1}}=k_{E^{c}_{3}}=2, \quad k_{E^{c}_{2}}=4\,.
\end{equation}
The resulting charged lepton and light neutrino mass matrices can be obtained from table~\ref{tab:sum_ch} and Eq.~\eqref{eq:nu_mass} respectively:
\begin{eqnarray}
\nonumber M_e&=&\begin{pmatrix}{\alpha Y^{(2)}_{\bm{3},1}}&{\alpha Y^{(2)}_{\bm{3},3}}&{\alpha Y^{(2)}_{\bm{3},2}}\\
	{\beta Y^{(4)}_{\bm{3},3}}&{\beta Y^{(4)}_{\bm{3},2}}&{\beta Y^{(4)}_{\bm{3},1}}\\
	{\gamma Y^{(2)}_{\bm{3},2}}&{\gamma Y^{(2)}_{\bm{3},1}}&{\gamma Y^{(2)}_{\bm{3},3}}\end{pmatrix}v\,,\\
M_{\nu}&=&\frac{{v}^2}{2\Lambda}\begin{pmatrix}{g_{1}Y^{(0)}_{\bm{1}}+{2g_{2}}Y^{(0)}_{\bm{3},1}}&{{-g_{2}}Y^{(0)}_{\bm{3},3}}&{{-g_{2}}Y^{(0)}_{\bm{3},2}}\\
	{{-g_{2}}Y^{(0)}_{\bm{3},3}}&{{2g_{2}}Y^{(0)}_{\bm{3},2}}&{g_{1}Y^{(0)}_{\bm{1}}-{g_{2}}Y^{(0)}_{\bm{3},1}}\\
	{{-g_{2}}Y^{(0)}_{\bm{3},2}}&{g_{1}Y^{(0)}_{\bm{1}}-{g_{2}}Y^{(0)}_{\bm{3},1}}&{{2g_{2}}Y^{(0)}_{\bm{3},3}}\end{pmatrix}\,,
\end{eqnarray}
where the  explicit matrix elements are determined by polyharmonic  Maa{\ss} forms of respective weights. This model can produce results consistent with experimental data for both NO and IO mass spectra, under conditions both with and without gCP symmetry. The predicted best fit values of input parameters and physical observable quantities are summarized in tables~\ref{tab:WO_NO_bf_noGCP}, \ref{tab:WO_NO_bf_GCP}, \ref{tab:WO_IO_bf_noGCP} and \ref{tab:WO_IO_bf_GCP}. After performing exhaustive scans of each model's parameter space while requiring all observables to remain within their experimental 3$\sigma$ ranges from NuFIT~\cite{Esteban:2024eli}, we find some interesting correlations between input parameters and observables. The correlations between input parameters and observables for the NO and IO mass spectra are shown in figures~\ref{fig:WO_mixing_NO} and~\ref{fig:WO_mixing_IO}, respectively, with green (red) indicating results without (with) gCP symmetry. Since gCP invariance requires all couplings to be real, CP violation can only arise from $\Re\tau$. Consequently, all CP-violating phases and input parameters are constrained to narrow intervals, as shown in figures~\ref{fig:WO_mixing_NO} and~\ref{fig:WO_mixing_IO}. Note that the three lepton mixing angles and the three CP violation phase parameters are constrained within narrow ranges for both mass hierarchies:
\begin{eqnarray}
\nonumber   \text{NO :} && \sin^2\theta_{13}\in[0.02030,0.02388] ([0.02030,0.02388]), \qquad \sin^2\theta_{12}\in[0.275,0.345] ([0.296,0.340]), \\
\nonumber &&  \sin^2\theta_{23} \in[0.501,0.506] ([0.501,0.506]), \qquad \delta_{CP}/\pi\in[1.211, 1.244] ([1.211, 1.242]), \\
\nonumber &&      \alpha_{21}/\pi\in[0.368, 1.419] ([0.883, 0.908]), \qquad \alpha_{31}/\pi\in[-0.451,0.342] ([-0.0628, -0.0455]),  \\
\nonumber &&  \sum_{i=1}^{3} m_{i}\in [94.20\,\text{meV},120\,\text{meV}] ( [94.29\,\text{meV},99.61\,\text{meV}]), \\
\nonumber &&  m_{\beta\beta}\in [5.673\,\text{meV},24.36\,\text{meV}] ( [6.068\,\text{meV},8.417\,\text{meV}])\,, \\
\nonumber \text{IO :} && \sin^2\theta_{13}\in[0.0212,0.0234] ([0.0212,0.0234]), \qquad \sin^2\theta_{12}\in[0.301,330] ([0.301,330]), \\ 
\nonumber &&  \sin^2\theta_{23} \in[0.544,0.555] ([0.544,0.554]), \qquad  \delta_{CP}/\pi\in[1.301,1.311] ([1.301,1.311]),     \\
\nonumber && \alpha_{21}/\pi\in[0.524,0.891] ([0.692, 0.719]), \qquad \alpha_{31}/\pi\in[0.566,1.066] ([0.811, 0.823]),  \\
\nonumber &&  \sum_{i=1}^{3} m_{i}\in [105.8\,\text{meV},120\,\text{meV}] ( [105.8\,\text{meV},110.8\,\text{meV}]), \\
&&  m_{\beta\beta}\in [20.56\,\text{meV},36.85\,\text{meV}] ( [26.29\,\text{meV},28.15\,\text{meV}])\,,
\end{eqnarray}
for model $\{C^{(2,4,2)}_{10},W_{3}\}$ without (with) gCP symmetry. It is remarkable that the atmospheric mixing angle $\theta_{23}$ is very closed to its maximal value for NO case. 
When gCP symmetry compatible with the $A_4$ modular symmetry is imposed, the Majorana CP phases $\alpha_{21}$ and $\alpha_{31}$ deviate slightly from their trivial values. These precise constraints imply that the model can be tested experimentally at upcoming long baseline neutrino facilities DUNE and T2HK which are highly sensitive to Dirac CP-violating phase and atmospheric mixing angle. Additionally, a strong correlation exists between the atmospheric angle $\theta_{23}$ and the Dirac CP phase $\delta_{CP}$ with the solar angle $\theta_{12}$. Such correlations could be probed by combining data from the future JUNO experiment with that from DUNE or T2HK. The neutrino mass parameters $\sum_{i=1}^{3} m_{i}$ and $m_{\beta\beta}$ are constrained across both mass orderings. For the NO case, $\sum_{i=1}^{3} m_{i}$ falls in the sensitivity reach of upcoming cosmological surveys such as Euclid+CMB-S4+LiteBIRD, while $m_{\beta\beta}$ lies below the current KamLAND-Zen limit but could be accessible to next-generation experiments like LEGEND-1000 and nEXO. In contrast, for the IO case, both parameters lie within projected detection ranges of future experiments, offering a clear avenue to distinguish between mass orderings.

\section{\label{sec:conclusion} Conclusion}

The modular invariance is a promising framework to describe the masses and mixing parameters of both quarks and leptons~\cite{Feruglio:2017spp}. In the framework of original modular symmetry, SUSY is a necessary component. In this context, the principle of modular invariance forces the Yukawa couplings to be level  $N$ modular forms which are holomorphic functions of the complex modulus $\tau$. However, there is no evidence for low-energy supersymmetry. Subsequently, a non-supersymmetric formulation of the modular flavor symmetry was recently proposed in Refs.~\cite{Qu:2024rns,Qu:2025ddz}. This approach extends the standard level $N$ modular forms by polyharmonic Maa{\ss} forms, a class of non-holomorphic modular forms that exist for zero, negative and positive integer weights. The framework of non-holomorphic modular flavor symmetry presents a novel avenue  for understanding the flavor structure of fermions.

In the present work, we perform a systematic analysis of all minimal lepton models based on non-holomorphic $\Gamma_3 \cong A_4$ modular symmetry. These models are explicitly constructed using only the modulus $\tau$, with no additional flavons. In these models, light neutrino masses are generated by the effective Weinberg operator, while the Yukawa couplings are derived from polyharmonic Maa{\ss} forms of level 3 with even weights $k$ in the range $-4 \leq k \leq 4$. 
The three LH lepton doublets transform as the $A_4$ triplet $\bm{3}$, while the RH charged leptons $E^c_{1,2,3}$ are assigned to $A_4$ singlets $\bm{1}$, $\bm{1^{\prime}}$ or $\bm{1^{\prime\prime}}$. 
According to the representation and weight assignments for the lepton fields, we identified 1820 independent minimal models, each depending on eight real parameters: the six dimensionless inputs in Eq.~\eqref{eq:six_inputs} and two overall scales. When the non-holomorphic $A_{4}$ modular flavor symmetry is extended to combine with gCP symmetry, then one more free parameter would be reduced for all these models. By numerically minimizing the $\chi^{2}$ function for each model, we find out 147 (6) phenomenologically viable models for the NO (IO) mass spectrum.  Among these, only 47 (5) remain consistent with experimental data from the lepton sector when gCP is imposed.  The corresponding best fit values of input parameters, lepton masses, lepton mixing angles, CP-violating phases, the 0$\nu\beta\beta$-decay effective Majorana mass and the kinematical mass are comprehensively presented in tables~\ref{tab:WO_NO_bf_noGCP}, \ref{tab:WO_NO_bf_GCP}, \ref{tab:WO_IO_bf_noGCP} and \ref{tab:WO_IO_bf_GCP}. 
Based on our predictions, the current JUNO constraint on $\sin^{2}\theta_{12}$ rules out only 5 of the 147 viable NO models without gCP symmetry, while all others remain consistent with experimental data. The future experimental results can significantly constrain the set of currently viable models. If the current best-fit value of $\sin^{2}\theta_{12}$ remains unchanged, JUNO will determine this parameter with high precision after six years of data. However, its projected $3\sigma$ interval still covers the $\sin^{2}\theta_{12}$ predictions of almost all models considered. The next-generation long baseline experiments like DUNE and T2HK are projected to determine the atmospheric mixing angle $\theta_{23}$ with unprecedented precision, which will rule out a large fraction of models.  Additional constraints are expected from refined measurements of the Dirac CP-violating phase $\delta_{CP}$, the sum of neutrino masses $\sum_{i=1}^{3}m_{i}$ and the effective Majorana mass $m_{\beta\beta}$ of 0$\nu\beta\beta$-decay. 
 
To illustrate our findings more clearly, we present detailed numerical results for the example model $\{C^{(2,4,2)}_{10}, W_{3}\}$, demonstrating how the non-holomorphic $A_{4}$ modular flavor symmetry can be applied to address the flavor problem. We present complete predictions for lepton mixing parameters, neutrino masses, and the effective mass of $0\nu\beta\beta$-decay under both NO and IO spectra, with and without gCP symmetry. Our analysis reveals several non-trivial correlations between input parameters and physical observables for both mass orderings. Furthermore, imposing gCP symmetry markedly reduces the allowed parameter space and yields much sharper, more definitive correlations. All these theoretical predictions can be rigorously tested by next-generation neutrino experiments. Most notably, the specific predictions of this model will face decisive verification from next-generation  $0\nu\beta\beta$-decay experiments such as LEGEND-1000 and nEXO.

\section*{Acknowledgements}

This work is supported by Natural Science Basic Research Program of Shaanxi (Program No. 2024JC-YBQN-0004), and the National Natural Science Foundation of China under Grant No. 12247103. 
 
\newpage

\begin{center}
	\renewcommand{\tabcolsep}{0mm}
	\renewcommand{\arraystretch}{1.15}
	\begin{small}
		\setlength\LTcapwidth{\textwidth}
		\setlength\LTleft{-0.0in}
		\setlength\LTright{0pt}
\begin{longtable}{|c|c|c|c|c|c|c|c|c|c|c|c|c|c|c|c|c|c|c|c|c|c|c|c|c|c|c|c|c|c|c|c|c|c|c|c|c|c|c|c|c|c|c|c|c|c|c|c|c|c|c|c|c|c|c|c|c|c|c|c|c|c|c|c|c|c|c|c|c|c|c|c|c|c|c|c|c|c|c|c|c|c|c|c|c|c|c|c|c|c|c|c|c|c|c|c|c|c|c|c|c|}
		\caption{\label{tab:WO_NO_bf_noGCP}
		The best fit values of the input parameters, three mixing angles $\theta_{12}$, $\theta_{13}$, $\theta_{23}$, Dirac CP-violating phase $\delta_{CP}$, Majorana CP-violating phases $\alpha_{21}$, $\alpha_{31}$,  three light neutrino masses $m_{1,2,3}$,  the effective mass $m_{\beta\beta}$ in $0\nu\beta\beta$-decay and the kinematical mass $m_{\beta}$ in beta decay  at the minimum values of $\chi^2$ for all viable models $\{C^{(k_{1},k_{2},k_{3})}_{m},W_{j}\}$ in the case of without gCP symmetry. The best fit values of the charged lepton masses are the global best fit values of them. The best fit values of the mass sum $\sum_{i=1}^{3} m_{i}$ and $\Delta m^{2}_{21}/\Delta m^{2}_{31}$ can be easy obtained from the three neutrino masses. In the following tables, we would not show them as the same reason.} \\
	\midrule
	\specialrule{0em}{1.0pt}{1.0pt}
			
	\endfirsthead
			
	\multicolumn{100}{c}
		{{\bfseries \tablename\ \thetable{} -- continued from previous page}} \\
		\hline

		\endhead

	\caption[]{continues on next page}\\
	\endfoot
			
	\endlastfoot
			
	\hline
	\texttt{Models} & \multicolumn{99}{c|}{ Best fit results for 148 viable models $\{C^{(k_{1},k_{2},k_{3})}_{m},W_{j}\}$ without gCP}   \\ \hline
	   $\{C^{(k_{1},k_{2},k_{3})}_{m},W_{j}\}$	&  \multicolumn{11}{c|}{  $\Re\langle \tau \rangle$} &  \multicolumn{11}{c|}{  $\Im\langle \tau \rangle$}  &  \multicolumn{11}{c|}{  $\beta/\alpha$} &  \multicolumn{11}{c|}{  $\gamma/\alpha$} &  \multicolumn{11}{c|}{  $|g_{2}/g_{1}|$} &  \multicolumn{11}{c|}{  $\text{arg}(g_{2}/g_{1})/\pi$} &  \multicolumn{11}{c|}{  $\alpha v$/MeV} & \multicolumn{11}{c|}{  $(g_{1}v^2/\Lambda)$/meV} &  \multicolumn{11}{c|}{$\chi^2_{\mathrm{min}}$} \\ \hline
			
	$\{C^{(-4,-2,0)}_{1},W_{1}\}$  & \multicolumn{11}{c|}{ $-$0.1647 } & \multicolumn{11}{c|}{ 1.092 } & \multicolumn{11}{c|}{ 35.10 } & \multicolumn{11}{c|}{ 836.8 } & \multicolumn{11}{c|}{ 0.4475 } & \multicolumn{11}{c|}{ 0.2843 } & \multicolumn{11}{c|}{ 5.803 } & \multicolumn{11}{c|}{ 910.7 } & \multicolumn{11}{c|}{ 8.396 }\\ \hline
	$\{C^{(-4,-2,4)}_{1},W_{1}\}$  & \multicolumn{11}{c|}{ 0.1311 } & \multicolumn{11}{c|}{ 1.131 } & \multicolumn{11}{c|}{ 0.004638 } & \multicolumn{11}{c|}{ 2.523 } & \multicolumn{11}{c|}{ 0.6208 } & \multicolumn{11}{c|}{ 1.735 } & \multicolumn{11}{c|}{ 593.8 } & \multicolumn{11}{c|}{ 637.4 } & \multicolumn{11}{c|}{ 3.255 }\\ \hline
	$\{C^{(-4,0,4)}_{1},W_{1}\}$  & \multicolumn{11}{c|}{ $-$0.1312 } & \multicolumn{11}{c|}{ 1.131 } & \multicolumn{11}{c|}{ 0.01677 } & \multicolumn{11}{c|}{ 2.523 } & \multicolumn{11}{c|}{ 0.6213 } & \multicolumn{11}{c|}{ 0.2651 } & \multicolumn{11}{c|}{ 593.8 } & \multicolumn{11}{c|}{ 637.5 } & \multicolumn{11}{c|}{ 0.1485 }\\ \hline
	$\{C^{(-4,2,4)}_{1},W_{1}\}$  & \multicolumn{11}{c|}{ 0.1311 } & \multicolumn{11}{c|}{ 1.131 } & \multicolumn{11}{c|}{ 0.002697 } & \multicolumn{11}{c|}{ 2.523 } & \multicolumn{11}{c|}{ 0.6207 } & \multicolumn{11}{c|}{ 1.735 } & \multicolumn{11}{c|}{ 593.9 } & \multicolumn{11}{c|}{ 637.5 } & \multicolumn{11}{c|}{ 3.253 }\\
\end{longtable}
\end{small}
\end{center}

\begin{center}
	\renewcommand{\tabcolsep}{0mm}
	\renewcommand{\arraystretch}{1.15}
	\begin{small}
		\setlength\LTcapwidth{\textwidth}
		\setlength\LTleft{0.0in}
		\setlength\LTright{0pt}
		\begin{longtable}{|c|c|c|c|c|c|c|c|c|c|c|c|c|c|c|c|c|c|c|c|c|c|c|c|c|c|c|c|c|c|c|c|c|c|c|c|c|c|c|c|c|c|c|c|c|c|c|c|c|c|c|c|c|c|c|c|c|c|c|c|c|c|c|c|c|c|c|c|c|c|c|c|c|c|c|c|c|c|c|c|c|c|c|c|c|c|c|c|c|c|}
		\caption{\label{tab:WO_NO_bf_GCP}
			The best-fit values of the input parameters, neutrino masses, and mixing parameters at the minimum $\chi^2$ for viable models $\{C^{(k_{1},k_{2},k_{3})}_{m},W_{j}\}$ with gCP symmetry. } \\
	\midrule
	\specialrule{0em}{1.0pt}{1.0pt}
				
	\endfirsthead
			
	\multicolumn{89}{c}
		{{\bfseries \tablename\ \thetable{} -- continued from previous page}} \\
\hline

	\endhead
				
	\caption[]{continues on next page}\\
	\endfoot
				
 \endlastfoot
				
	\hline

	\texttt{Models} &\multicolumn{89}{|c|}{Best fit results for 47 viable models $\{C^{(k_{1},k_{2},k_{3})}_{m},W_{j}\}$ with gCP} \\ \hline
	$\{C^{(k_{1},k_{2},k_{3})}_{m},W_{j}\}$ & \multicolumn{11}{c|}{$\Re\langle \tau \rangle$} & \multicolumn{11}{c|}{$\Im\langle \tau \rangle$}  & \multicolumn{11}{c|}{$\beta/\alpha$} & \multicolumn{11}{c|}{$\gamma/\alpha$} & \multicolumn{11}{c|}{$g_{2}/g_{1}$} & \multicolumn{11}{c|}{$\alpha v$/MeV} & \multicolumn{11}{c|}{$(g_{1}v^2/\Lambda)/$meV} &  \multicolumn{11}{c|}{$\chi^2_{\mathrm{min}}$}\\ \hline
	$\{C^{(-4,-2,2)}_{2},W_{3}\}$  & \multicolumn{11}{c|}{ $-$0.1123 } & \multicolumn{11}{c|}{ 1.107 } & \multicolumn{11}{c|}{ 1532.0 } & \multicolumn{11}{c|}{ 100.4 } & \multicolumn{11}{c|}{ $-$3.292 } & \multicolumn{11}{c|}{ 1.925 } & \multicolumn{11}{c|}{ 175.4 } & \multicolumn{11}{c|}{ 5.904 }\\ \hline
	$\{C^{(-4,-2,2)}_{3},W_{1}\}$ & \multicolumn{11}{c|}{ $-$0.01204 } & \multicolumn{11}{c|}{ 1.082 } & \multicolumn{11}{c|}{ 248.3 } & \multicolumn{11}{c|}{ 882.9 } & \multicolumn{11}{c|}{ $-$1.588 } & \multicolumn{11}{c|}{ 1.707 } & \multicolumn{11}{c|}{ 418.6 } & \multicolumn{11}{c|}{ 2.010 }\\ \hline
	$\{C^{(-4,-2,-2)}_{4},W_{3}\}$ & \multicolumn{11}{c|}{ 0.3203 } & \multicolumn{11}{c|}{ 0.9639 } & \multicolumn{11}{c|}{ 65.85 } & \multicolumn{11}{c|}{ 1065.0 } & \multicolumn{11}{c|}{ 3.867 } & \multicolumn{11}{c|}{ 3.527 } & \multicolumn{11}{c|}{ 186.9 } & \multicolumn{11}{c|}{ 12.20 }\\ \hline
	$\{C^{(-4,2,0)}_{4},W_{1}\}$ & \multicolumn{11}{c|}{ $-$0.3678 } & \multicolumn{11}{c|}{ 0.9781 } & \multicolumn{11}{c|}{ 7.439 } & \multicolumn{11}{c|}{ 535.0 } & \multicolumn{11}{c|}{ 1.583 } & \multicolumn{11}{c|}{ 11.41 } & \multicolumn{11}{c|}{ 607.5 } & \multicolumn{11}{c|}{ 25.69 }\\
\end{longtable}
	\end{small}
	\end{center}
	
\begin{center}
	\renewcommand{\tabcolsep}{0.3mm}
	\renewcommand{\arraystretch}{1.15}
	\begin{small}
		\setlength\LTcapwidth{\textwidth}
		\setlength\LTleft{-0.0in}
		\setlength\LTright{0pt}
		\begin{longtable}{|c|c|c|c|c|c|c|c|c|c|c|c|c|c|c|c|c|c|c|c|c|c|c|c|c|c|c|c|c|c|c|c|c|c|c|c|c|c|c|c|c|c|c|c|c|c|c|c|c|c|c|c|c|c|c|c|c|c|c|c|c|c|c|c|c|c|c|c|c|c|c|c|c|c|c|c|c|c|c|c|c|c|c|c|c|c|c|c|c|c|c|c|c|c|c|c|c|c|c|c|c|c|}
		\caption{\label{tab:WO_IO_bf_noGCP}
		The best-fit values of the input parameters, neutrino masses, and mixing parameters at the minimum $\chi^2$ for viable models $\{C^{(k_{1},k_{2},k_{3})}_{m},W_{j}\}$ without gCP symmetry.} \\
	\midrule
	\specialrule{0em}{1.0pt}{1.0pt}
			
	\endfirsthead
			
	\multicolumn{100}{c}
	{{\bfseries \tablename\ \thetable{} -- continued from previous page}} \\
\hline

	\endhead

	\caption[]{continues on next page}\\
	\endfoot
			
\endlastfoot
			
	\hline
	\texttt{Models} & \multicolumn{99}{c|}{ Best fit results for 6 viable models $\{C^{(k_{1},k_{2},k_{3})}_{m},W_{j}\}$ without gCP}   \\ \hline
	   $\{C^{(k_{1},k_{2},k_{3})}_{m},W_{j}\}$	&  \multicolumn{11}{c|}{  $\Re\langle \tau \rangle$} &  \multicolumn{11}{c|}{  $\Im\langle \tau \rangle$}  &  \multicolumn{11}{c|}{  $\beta/\alpha$} &  \multicolumn{11}{c|}{  $\gamma/\alpha$} &  \multicolumn{11}{c|}{  $|g_{2}/g_{1}|$} &  \multicolumn{11}{c|}{  $\text{arg}(g_{2}/g_{1})/\pi$} &  \multicolumn{11}{c|}{  $\alpha v$/MeV} & \multicolumn{11}{c|}{  $(g_{1}v^2/\Lambda)$/meV} &  \multicolumn{11}{c|}{$\chi^2_{\mathrm{min}}$} \\ \hline
	 $\{C^{(-4,-2,-4)}_{7},W_{3}\}$  & \multicolumn{11}{c|}{ $-$0.418 } & \multicolumn{11}{c|}{ 1.05 } & \multicolumn{11}{c|}{1794.0 } & \multicolumn{11}{c|}{ 186.6 } & \multicolumn{11}{c|}{ 42.8 } & \multicolumn{11}{c|}{ 1.57 } & \multicolumn{11}{c|}{ 1.74 } & \multicolumn{11}{c|}{ 20.5 } & \multicolumn{11}{c|}{ 8.27 }\\ \hline
	$\{C^{(-4,-2,2)}_{7},W_{3}\}$  & \multicolumn{11}{c|}{ 0.459 } & \multicolumn{11}{c|}{ 1.06 } & \multicolumn{11}{c|}{ 108.0 } & \multicolumn{11}{c|}{ 843.3 } & \multicolumn{11}{c|}{ 19.5 } & \multicolumn{11}{c|}{ 0.523 } & \multicolumn{11}{c|}{ 1.67 } & \multicolumn{11}{c|}{ 44.1 } & \multicolumn{11}{c|}{ 12.2 }\\ \hline
	 $\{C^{(-2,2,-4)}_{7},W_{3}\}$  & \multicolumn{11}{c|}{ $-$0.418 } & \multicolumn{11}{c|}{ 1.05 } & \multicolumn{11}{c|}{ 0.000155 } & \multicolumn{11}{c|}{ 0.104 } & \multicolumn{11}{c|}{ 13.3 } & \multicolumn{11}{c|}{ 1.52 } & \multicolumn{11}{c|}{ 3123.0 } & \multicolumn{11}{c|}{ 65.8 } & \multicolumn{11}{c|}{ 8.27 }\\ \hline
	 $\{C^{(-4,2,-2)}_{9},W_{3}\}$  & \multicolumn{11}{c|}{ $-$0.418 } & \multicolumn{11}{c|}{ 1.05 } & \multicolumn{11}{c|}{ 0.00591 } & \multicolumn{11}{c|}{ 9.62 } & \multicolumn{11}{c|}{ 32.4 } & \multicolumn{11}{c|}{ 0.447 } & \multicolumn{11}{c|}{ 324.7 } & \multicolumn{11}{c|}{ 27.0 } & \multicolumn{11}{c|}{ 8.29 }\\
\end{longtable}
\end{small}
\end{center}

\begin{center}
\renewcommand{\tabcolsep}{0.3mm}
\renewcommand{\arraystretch}{1.15}
\begin{small}
\setlength\LTcapwidth{\textwidth}
\setlength\LTleft{0.0in}
\setlength\LTright{0pt}
\begin{longtable}{|c|c|c|c|c|c|c|c|c|c|c|c|c|c|c|c|c|c|c|c|c|c|c|c|c|c|c|c|c|c|c|c|c|c|c|c|c|c|c|c|c|c|c|c|c|c|c|c|c|c|c|c|c|c|c|c|c|c|c|c|c|c|c|c|c|c|c|c|c|c|c|c|c|c|c|c|c|c|c|c|c|c|c|c|c|c|c|c|c|c|}
\caption{\label{tab:WO_IO_bf_GCP}
	The best-fit values of the input parameters, neutrino masses, and mixing parameters at the minimum $\chi^2$ for viable models $\{C^{(k_{1},k_{2},k_{3})}_{m},W_{j}\}$ with gCP symmetry. } \\
	\midrule
\specialrule{0em}{1.0pt}{1.0pt}
			
\endfirsthead
			
\multicolumn{89}{c}
	{{\bfseries \tablename\ \thetable{} -- continued from previous page}} \\
\hline

\endhead
			
\caption[]{continues on next page}\\
\endfoot
			
\endlastfoot
			
\hline

     \texttt{Models} &\multicolumn{88}{|c|}{Best fit results for 5 viable models $\{C^{(k_{1},k_{2},k_{3})}_{m},W_{j}\}$ with gCP} \\ \hline
     $\{C^{(k_{1},k_{2},k_{3})}_{m},W_{j}\}$ & \multicolumn{11}{c|}{$\Re\langle \tau \rangle$} & \multicolumn{11}{c|}{$\Im\langle \tau \rangle$}  & \multicolumn{11}{c|}{$\beta/\alpha$} & \multicolumn{11}{c|}{$\gamma/\alpha$} & \multicolumn{11}{c|}{$g_{2}/g_{1}$} & \multicolumn{11}{c|}{$\alpha v$/MeV} & \multicolumn{11}{c|}{$(g_{1}v^2/\Lambda)/$meV} &  \multicolumn{11}{c|}{$\chi^2_{\mathrm{min}}$}\\ \hline
   $\{C^{(-4,-2,-4)}_{7},W_{3}\}$  & \multicolumn{11}{c|}{ $-$0.418 } & \multicolumn{11}{c|}{ 1.05 } & \multicolumn{11}{c|}{ 1794.0 } & \multicolumn{11}{c|}{ 186.6 } & \multicolumn{11}{c|}{ 197.1 } & \multicolumn{11}{c|}{ 1.74 } & \multicolumn{11}{c|}{ 4.44 } &  \multicolumn{11}{c|}{ 8.27 }\\ \hline
   $\{C^{(-4,-2,2)}_{7},W_{3}\}$  & \multicolumn{11}{c|}{ 0.459 } & \multicolumn{11}{c|}{ 1.06 } & \multicolumn{11}{c|}{ 108.0 } & \multicolumn{11}{c|}{ 843.3 } & \multicolumn{11}{c|}{ $-$266.3 } & \multicolumn{11}{c|}{ 1.67 } & \multicolumn{11}{c|}{ 3.23 } & \multicolumn{11}{c|}{ 12.2 }\\ \hline
   $\{C^{(-4,2,-2)}_{9},W_{3}\}$  & \multicolumn{11}{c|}{ $-$0.418 } & \multicolumn{11}{c|}{ 1.05 } & \multicolumn{11}{c|}{ 0.00591 } & \multicolumn{11}{c|}{ 9.62 } & \multicolumn{11}{c|}{ 197.3 } & \multicolumn{11}{c|}{ 324.7 } & \multicolumn{11}{c|}{ 4.44 } &   \multicolumn{11}{c|}{ 8.29 }\\ \hline
   $\{C^{(2,4,-2)}_{9},W_{3}\}$  & \multicolumn{11}{c|}{ $-$0.459 } & \multicolumn{11}{c|}{ 1.05 } & \multicolumn{11}{c|}{ 0.00428 } & \multicolumn{11}{c|}{ 37.5 } & \multicolumn{11}{c|}{ 10000.0 } & \multicolumn{11}{c|}{ 82.6 } & \multicolumn{11}{c|}{ 0.0869 } &  \multicolumn{11}{c|}{ 26.3 }\\
\end{longtable}
	\end{small}
\end{center}

\providecommand{\href}[2]{#2}\begingroup\raggedright\endgroup

\end{document}